%% file: AICM.tex
\documentclass[sigconf]{acmart}

\AtBeginDocument{%
  \providecommand\BibTeX{{%
    \normalfont B\kern-0.5em{\scshape i\kern-0.25em b}\kern-0.8em\TeX}}}
    
\newtheorem{theorem}{Theorem}[section]

\newtheorem{definition}{Definition}[section]

\setcopyright{iw3c2w3}

\copyrightyear{2021}
\acmYear{2021}
\setcopyright{iw3c2w3}
\acmConference[WWW '21]{Proceedings of the Web Conference 2021}{April 19--23, 2021}{Ljubljana, Slovenia}
\acmBooktitle{Proceedings of the Web Conference 2021 (WWW '21), April 19--23, 2021, Ljubljana, Slovenia}
\acmPrice{}
\acmDOI{10.1145/3442381.3449913}
\acmISBN{978-1-4503-8312-7/21/04}

\usepackage{multirow}
\usepackage{subfigure}
\usepackage{enumitem}
\usepackage{enumerate}\allowdisplaybreaks

\newcommand{\minisection}[1]{\vspace{3pt}\noindent\textbf{#1.}}

\begin{document}

\title{An Adversarial Imitation Click Model \\for Information Retrieval}

\author{Xinyi Dai$^1$, Jianghao Lin$^1$, Weinan Zhang$^1$, Shuai Li$^1$, Weiwen Liu$^2$}
\author{Ruiming Tang$^2$, Xiuqiang He$^2$, Jianye Hao$^2$, Jun Wang$^3$, Yong Yu$^1$}
\affiliation{$^1$Shanghai Jiao Tong University, $^2$Huawei Noah's Ark Lab, $^3$University College London}
\email{{daixinyi, chiangel, wnzhang, 	shuaili8, yyu}@sjtu.edu.cn, {liuweiwen8, tangruiming, hexiuqiang1, haojianye}@huawei.com}
\email{jun.wang@cs.ucl.ac.uk}


\renewcommand{\shortauthors}{X. Dai, et al.}
\renewcommand{\shorttitle}{AICM}
\settopmatter{printacmref=false}
\begin{abstract}
    Modern information retrieval systems, including web search, ads placement, and recommender systems, typically rely on learning from user feedback. Click models, which study how users interact with a ranked list of items, provide a useful understanding of user feedback for learning ranking models. Constructing "right" dependencies is the key of any successful click model. However, probabilistic graphical models (PGMs) have to rely on manually assigned dependencies, and oversimplify user behaviors. 
    Existing neural network based methods promote PGMs by enhancing the expressive ability and allowing flexible dependencies, but still suffer from exposure bias and inferior estimation. In this paper, we propose a novel framework, Adversarial Imitation Click Model (AICM), based on imitation learning. Firstly, we explicitly learn the reward function that recovers users' intrinsic utility and underlying intentions. Secondly, we model user interactions with a ranked list as a dynamic system instead of one-step click prediction, alleviating the exposure bias problem. Finally, we minimize the JS divergence through adversarial training and learn a stable distribution of click sequences, which makes AICM generalize well across different distributions of ranked lists. 
    A theoretical analysis has indicated that AICM reduces the exposure bias from $O(T^2)$ to $O(T)$. Our studies on a public web search dataset show that AICM not only outperforms state-of-the-art models in traditional click metrics but also achieves superior performance in addressing the exposure bias and recovering the underlying patterns of click sequences. 
\end{abstract}




\maketitle

\input{text/intro.tex}
\input{text/method.tex}
\input{text/exp.tex}
\input{text/related.tex}

\section*{Acknowledgement}
    We thank Minghuan Liu and Jian Shen for helpful discussions.
    The corresponding author Weinan Zhang is supported by ``New Generation of AI 2030'' Major Project (2018AAA0100900) and National Natural Science Foundation of China (62076161, 61772333, 61632017). The work is also sponsored by Huawei Innovation Research Program.

\bibliographystyle{ACM-Reference-Format}
\bibliography{AICM}

\clearpage
\appendix
\input{text/Appendix}

\end{document}

%% file: text/intro.tex
\section{Introduction}

Learning ranking functions from user behaviors (such as click logs) is a critical task in web search~\cite{wu2018turning}, ads placement~\cite{joachims2016counterfactual}, and recommender systems~\cite{zhao2019recommending}. 
To better understand user behaviors and derive a ranking function that best fulfills users' information needs, various click models have been developed~\cite{dupret2008user,srikant2010user,borisov2016neural,chen2020context}. Click models characterize how users interact with a list of items. Given click logs (including a set of queries, a ranked list of items, and the click data for each query), click models are trained to predict a sequence of user clicks, and return a set of model parameters that reflect users' underlying behaviors~\cite{chuklin2015click}. Click models provide useful evidence for ranking functions in both training and testing processes. In training, click models generate users' feedback on items with specific positions and contexts that have not been seen in the click logs, which help alleviate the inherent biases in users' behaviors (e.g., position bias, presentation bias)~\cite{joachims2017accurately,yue2010beyond}. In testing, click models can be applied to evaluating the performance of ranking functions in cases where real users are not available or negative impacts on user experience have to be avoided. 

Earlier click models are based on the probabilistic graphic models (PGMs). They represent user behaviors as a sequence of observable and hidden states, e.g., clicks, skips, attractiveness, and examinations~\cite{borisov2016neural}. Each state is defined as a binary event, e.g., whether a user examines a document, or whether a user is attracted by a document. Yet PGM framework requires manually setting the dependencies between the events, and thereby may be over-simplified and overlook some key aspects in user behaviors. Moreover, the expressive ability of PGM framework is usually limited \cite{borisov2016neural,chen2020context}. 

To improve the expressive ability and allow flexible dependencies, \citet{borisov2016neural} proposed the neural click model (NCM). Rather than using binary random variables, NCM represents user behaviors as vector sequences with the \textit{distributed vector representation approach}. The click sequence model (CSM)~\cite{borisov2018click} and the context-aware click model (CACM)~\cite{chen2020context} utilize complex model structures to incorporate more information (e.g., session context information) and thus further enhance the expressive ability. However, such methods still suffer from several limitations. 

First, the ultimate goal of click models is to understand user behaviors, and most importantly, the intrinsic utility behind the behaviors, e.g., maximizing the information needs or minimizing the effort of information seeking. 
This is the underlying intention that the user performs certain actions like clicks and skips.
If this utility is modeled explicitly, then it will not only help click modeling, but also provide us with insights and quantitative guidance for the optimization and evaluation of a ranking function.
Simply treating click model as a click prediction task with a black-box neural network might ignore this important aspect of click model.

Second, existing neural network (NN)-based models generally overlook the problem of \textit{exposure bias} \cite{bengio2015scheduled}, which refers to the model input discrepancy between training and testing. Specifically, the neural network based methods mentioned above predict the next click based on previous clicks of the ground truth sequences in the training procedure. During testing, however, these models have to predict successive clicks based on previous predictions made by itself, which have not been seen during training. This discrepancy comes from the conflicts between the dynamic nature of user behaviors and the static modeling of these models. 
User behaviors naturally depend on the previously happened ones. However, previous works are supervised to predict one click at each time step by assuming all previous clicks are correct. Such a greedy method may yield sub-optimal results since small errors accumulated at each time step leads to a great deviation from the optimal sequence~\cite{ross2011reduction}. 

Moreover, existing NN-based models may not generalize well when the test data deviates from the training data, especially when the data is rather sparse w.r.t. to the whole space of click sequences. User behaviors are complex in nature and may contain multiple patterns. If the data is sparse, using maximum likelihood estimation (MLE) as in prior works~\cite{borisov2016neural,chen2020context} tends to average on all the patterns and thus fails to fit the complex user behaviors, resulting in \textit{inferior estimation}. The MLE objective function only minimizes the KL divergence between target distribution and learned distribution, i.e., the forward KL divergence. However, the KL divergence between learned distribution and target distribution, i.e., the reverse KL divergence, 
which concentrates on the major pattern~\cite{bishop2006pattern}, has the potential to be beneficial for the click model to achieve better performance under its own generated distribution.

To tackle the aforementioned limitations, in this work, we propose a novel learning paradigm for click models based on imitation learning framework, namely, Adversarial Imitation Click Model (AICM). Imitation learning is a learning paradigm that aims at reconstructing sequential decision-making policies from sampled experts' trajectories~\cite{ho2016generative}. 
\textit{Firstly}, we regard user behaviors as expert demonstrations and thus assume that users' intrinsic utility is maximized. With this assumption, we build a reward function explicitly from users' click logs. Then we use this reward function to guide the learning of a click policy that reproduces user behaviors. 
This reward function provides important insights and quantitative guidance for the optimization and evaluation of a ranking function. 
\textit{Secondly}, we formulate the click model as a dynamic system. To be specific, we base users' current state on previous predictions and optimize the click model for a long-term objective rather than a short sighted loss over individual clicks, which alleviates the exposure bias problem. \textit{Finally}, we solve the dynamic system via an imitation learning algorithm, more specifically the generative adversarial imitation learning (GAIL) algorithm, to minimize the Jensen-Shannon (JS) divergence between the distributions of target sequences and generated sequences. Instead of solely considering the forward KL divergence as in MLE, minimizing JS divergence helps to learn a more stable distribution of click sequences, which makes the click model generalize better on different distributions of ranked lists.

Our theoretical analysis shows that AICM reduces the exposure bias from $O(T^2)$ to $O(T)$.
Extensive empirical studies are conducted to show the state-of-the-art performance of AICM in traditional click prediction and relevance estimation tasks, and superior performance in addressing the exposure bias and recovering the underlying patterns of click sequences. 
The results also demonstrate that AICM generalizes well on different distributions of ranked lists and achieves stable performance even in bad cases, which allows safe exploration of the ranking functions.

%% file: text/method.tex
\section{Preliminary: Imitation Learning}
The goal of imitation learning is to learn a behavior policy $\pi_\theta(a|s)$ that reproduces expert behaviors, given a set of expert demonstrations, where each of such demonstrations is a sequence of states and actions, i.e., $\mathbf{\tau}_{\pi_E} = [s_0, a_0, s_1, a_1, \cdots]$~\cite{osa2018algorithmic}. Models of imitation learning are generally divided into three classes: behavior cloning (BC), inverse reinforcement learning (IRL) and generative adversarial imitation learning (GAIL).

\subsection{Behavior Cloning}
Behavior cloning (BC)~\cite{bain1995framework} learns a policy that directly maps states to actions without recovering the reward function. BC maximizes the likelihood of experts' trajectories, 
\begin{equation}
\mathop{\max}_\theta \mathbb{E}_{(s,a)\sim\tau_{\pi_E}} [\log {\pi_\theta(a|s)}].
\label{eq:bcobj}
\end{equation}
which equals to the minimization of KL divergence $D_{KL}(\pi_E, \pi_\theta)$ for each state visited by the expert policy. 

If we regard the learning of click models as an imitation learning problem, the traditional supervised click models can be categorized into BC. 
However, BC suffers from compounding error since it only fits a single-step decision instead of focusing on a long-horizon planning~\cite{ross2011reduction}. 

\subsection{Inverse Reinforcement Learning}
Inverse Reinforcement Learning (IRL) recovers the reward function from expert demonstrations under the assumption that such demonstrations are optimal. Then a policy can be trained according to the learned reward function. The IRL problem is ill-posed because a policy can be optimal for multiple reward functions. To obtain the unique solution, various additional objectives such as maximum margin~\cite{ng2000algorithms, abbeel2004apprenticeship} and maximum entropy~\cite{ziebart2008maximum, ziebart2010modeling} have been proposed. Taking the maximum causal entropy IRL~\cite{ziebart2008maximum, ziebart2010modeling} as an example, it looks for a cost function $c\in \mathcal{C}$ (where the cost is equivalent to negative reward) that assigns low cost to the expert policy $\pi_E$ and high cost to the other policies, 
\begin{equation}
\mathop{\max}_{c\in \mathcal{C}} \big\{ (\mathop{\min}_{\theta} \mathbb{E}_{\pi_\theta}[c(s, a)]-H(\pi_\theta)) - \mathbb{E}_{\pi_E}[c(s, a)] \big\},
\end{equation}
where $H(\pi_\theta) = \mathbb{E}_{\pi_\theta}[-\log \pi_\theta(a|s)]$ is the causal entropy of the learned policy $\pi_\theta$. 
We use an expectation w.r.t. policy $\pi$ to denote an expectation w.r.t. the trajectory it generates, e.g., $\mathbb{E}_\pi[c(s,a)] = \mathbb{E}_{s_0\sim p_0, a_t\sim\pi(\cdot|s_t), s_{t+1}\sim P(\cdot|s_{t}, a_{t})}[(\sum_{t=0}^T \gamma^t c(s_t,a_t))]$, where $\gamma$ is the discount factor. 
IRL methods often require a costly iterative learning process, which has to solve an RL-type problem in every update step of the reward function. 

\subsection{Generative Adversarial Imitation Learning}
Inspired by the connection between GANs~\cite{goodfellow2014generative} and IRL, 
\citet{ho2016generative} proposed generative adversarial imitation learning (GAIL). GAIL trains a policy $\pi_\theta(a|s)$ with the reward provided by a discriminator $D_w(s,a): \mathcal{S} \times \mathcal{A} \rightarrow (0,1)$, which distinguishes between state-action pairs of $\pi_{\theta}$ and $\pi_E$. The objective function of GAIL is:
\begin{equation}
    \mathop{\min}_{\theta} \mathop{\max}_{w} \mathbb{E}_{\pi_\theta}[\log (D_w(s,a))] + \mathbb{E}_{\pi_E}[\log (1 - D_w(s,a))] - \lambda H(\pi_\theta).
\label{eq:obj1}
\end{equation}

\citet{ho2016generative} showed that IRL is a dual of the occupancy measure matching under the maximum entropy principle. GAIL essentially solves an occupancy measure matching problem by minimizing the JS divergence between the occupancy measure under the learned behavior policy and the expert policy, with the causal entropy $H(\pi_\theta)$ as the policy regularizer:
\begin{equation}
    \mathop{\min}_{\theta} D_{JS}(\rho_{\pi_\theta}, \rho_{\pi_E}) - \lambda H(\pi_\theta).
\label{eq:obj2}
\end{equation}
where
\begin{equation}
D_{JS}(\rho_{\pi_\theta}, \rho_{\pi_E}) = D_{KL}\Big(\rho_{\pi_\theta}, \frac{\rho_{\pi_\theta} + \rho_{\pi_E}}{2}\Big) + D_{KL}\Big(\rho_{\pi_E}, \frac{\rho_{\pi_\theta} + \rho_{\pi_E}}{2}\Big).
\end{equation}

Here the normalized occupancy measure $\rho_{\pi_\theta}$ and $\rho_{\pi_E}$ denote
the distributions of state-action pairs under the learned behavior policy and the expert policy respectively. To be specific, $\rho_{\pi}(s,a) = \frac{1-\gamma}{1-\gamma^{T+1}}\sum_{t=0}^T \gamma^t P_\pi(s_t=s, a_t=a)$, where $\frac{1-\gamma}{1-\gamma^{T+1}}$ is a normalization to ensure the probability sum equals one.
As can be seen, JS divergence considers both forward and reverse KL divergence between the learned behavior policy and the target policy, making the learned behavior policy precise and stable. 

\section{Methodology}
We first formulate the click model as an imitation learning problem and then present the overview of our proposed method AICM. After that, we introduce each component of AICM in detail. 

\subsection{Problem Formulation}
\begin{figure}[h]
      \centering
      \includegraphics[width=0.46\textwidth]{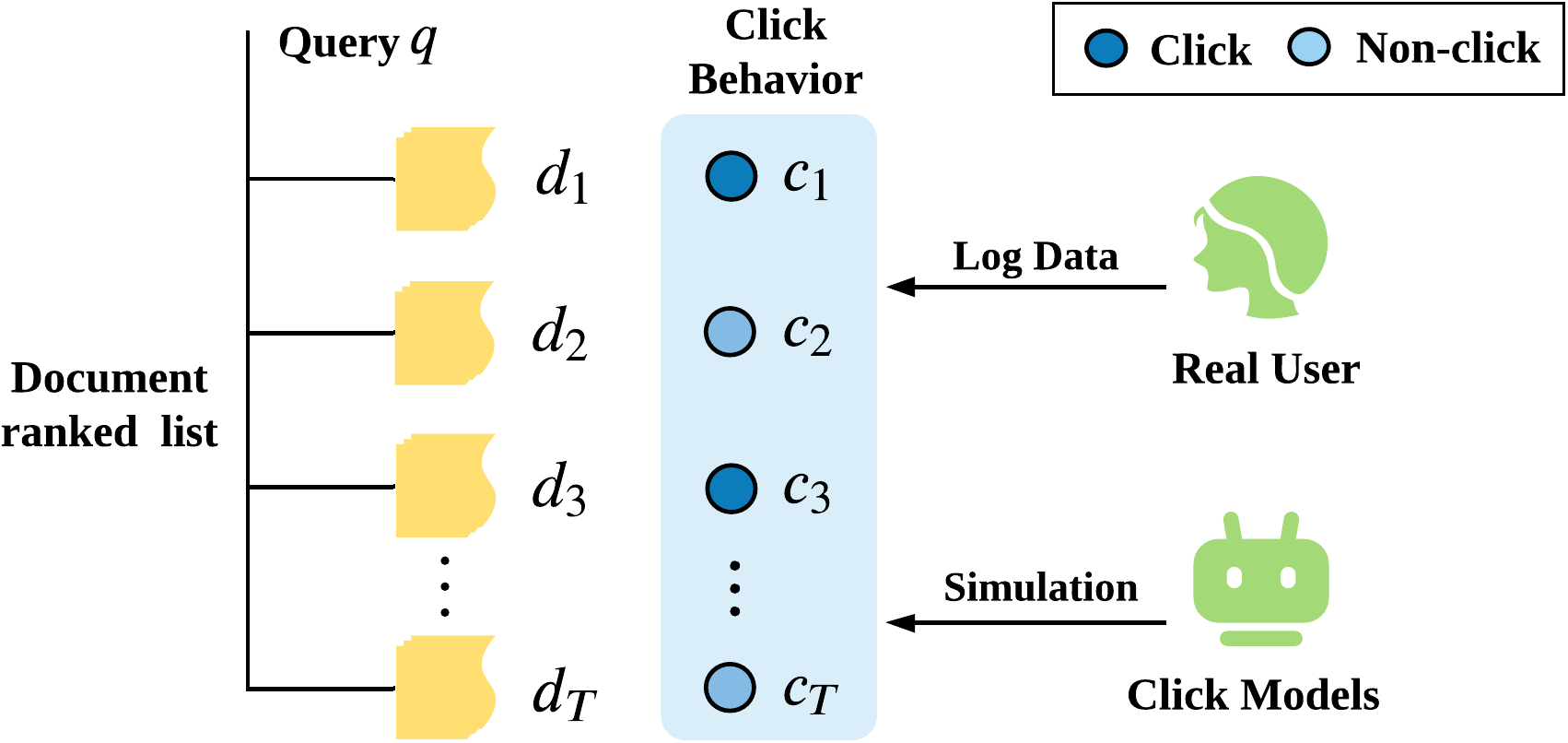} 
      \caption{An overview on click model}
      \label{fig:click_model}
\end{figure}

Users' interaction with a ranked list can be naturally interpreted as a sequential decision-making process. As illustrated in Figure~\ref{fig:click_model}, a user starts a search session by issuing a query $q$, and the ranking system delivers a ranked list with $T$ corresponding documents $D = \{d_{1}, d_{2}, \ldots, d_{T}\}$. The user examines the presented list, possibly clicks one or more documents, and then abandons the list to end the interaction. During such a process, a click sequence $\{c_1,\ldots,c_T\}$ is generated. The goal of click models~\cite{borisov2016neural} is to simulate the process from issuing a query till abandoning the search. 

Note that in this paper we consider a general click model setting, where the information of the user's previous search queries in the same session (like in CACM~\cite{chen2020context}) are not considered.

As a sequential decision-making process, a Markov decision process (MDP) can be used to model user behaviors, where the key components are defined as follows. 
\begin{itemize}[leftmargin=10pt]
    \item \textbf{State.} The initial user state $s_0$ is initialized with the query $q$. The state $s_{t}$ contains current document $d_t$ and the user's interactions $c_1,\ldots,c_{t-1}$ (click or not) with the documents before rank $t$, specifically $s_t =\{q, d_1, \ldots, d_{t-1}, d_t, c_1, \ldots, c_{t-1}\}$.
    \item \textbf{Action.} An action $a_t$ is the user's interaction with document $d_t$ at rank $t$, i.e., $a_t = c_t$. Whether the user clicks on an item is based on the policy $\pi_\theta (a_t|s_t)$, as parameterized by $\theta$. 
    \item \textbf{Transition.} The state $s_t$ updates according to the user's interaction $c_t$ and the next document $d_{t+1}$, i.e., $s_{t+1} = \{s_t, c_t, d_{t+1}\}$ for $t>0$. Note that when $t=0$, we have $s_1 = \{s_0, d_1\}$. 
\end{itemize}

Similar to many existing click models~\cite{borisov2016neural, chen2020context}, 
we do not model the abandonment explicitly, rather our model is trained to predict low click probabilities for documents that are unlikely to be examined by the user. The state at the end of the ranked list is simply set as the terminal state.

\subsection{Overview of AICM}
\begin{figure*}[t]
	\centering
	\includegraphics[width=0.9\textwidth]{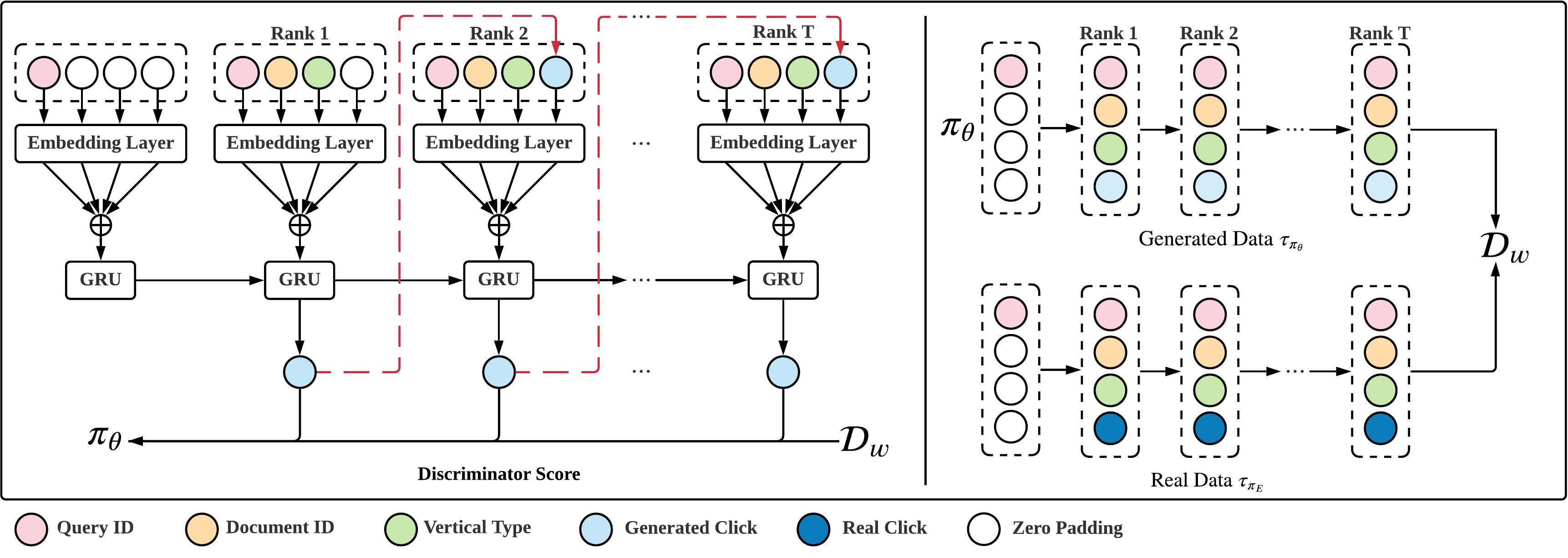}
    \caption{The overall framework of AICM. Left: model architecture of the generator. Right: model architecture of the discriminator. The blank node is zero padding which will be mapped to the zero vector of corresponding shape as described in Section~\ref{sec:generator} and Section~\ref{sec:discriminator} after the embedding layer.}
    \label{fig:overall framework}
\end{figure*}

In imitation learning, we aim to learn a behavior policy $\pi_\theta(a|s)$ from the state-action sequences provided by experts, i.e., expert demonstrations. 
In this work, we propose Adversarial Initation Click Model (AICM), by adopting GAIL framework to imitate the expert policy. AICM consists of three parts: 1) embedding layer for the query, document and interaction representations; 2) generator $\pi_\theta(a|s)$, i.e., the behavior policy, that generates user clicks; and 3) discriminator $D_w(s, a)$ that measures the difference between the generated user clicks and the ground truth clicks, as parameterized by $w$. To alleviate the exposure bias, the generator and discriminator are learned in an adversarial training paradigm. The overall framework for AICM is shown in Figure~\ref{fig:overall framework}. 

We describe how query $q$, document $d_{t}$ and interaction $c_{t}$ are represented in the embedding layer. 
For each document $d_{t}$, we incorporate its vertical type $v_{t}$, which is an important feature that infers the presentation style for each document. Common vertical types for a commercial search engines include the organic result, the encyclopedia vertical, the illustrated vertical, and etc. We first transform the original ID feature into a high-dimensional sparse features via one-hot encoding. Then the embedding layer is performed on the one-hot vectors to map them to low-dimensional, dense real-value embedding vectors:
\begin{equation}
\begin{aligned}
    &\mathbf{v}_{q}=\mathbf{Em b}_{\mathbf{q}}\left(q\right), 
    &\mathbf{v}_{d}=\mathbf{Em b}_{\mathbf{d}}\left(d_{t}\right),\\
    &\mathbf{v}_{v}=\mathbf{Em b}_{\mathbf{v}}\left(v_{t}\right), 
    &\mathbf{v}_{c}=\mathbf{Em b}_{\mathbf{c}}\left(c_{t}\right),
\end{aligned}
\end{equation}
where $\mathbf{Em b}_{\mathbf{q}} \in \mathbb{R}^{N_q \times l_q}$,  $\mathbf{Em b}_{\mathbf{d}} \in \mathbb{R}^{N_d \times l_d}$, $\mathbf{Em b}_{\mathbf{v}} \in \mathbb{R}^{N_v \times l_v}$,  $\mathbf{Em b}_{\mathbf{c}} \in \mathbb{R}^{N_c \times l_c}$, $N_*$ and $l_*$ denote the input size and the embedding size of each feature, respectively\footnote{When there is no ambiguity, we omit the subscripts of embeddings for the ease of presentation.}. 

\subsection{Generator}
\label{sec:generator}
The generator $\pi_\theta (a|s)$ generates users' feedback based on state $s$. The state $s$ carries the information of users' historical interactions with the documents presented before. We mainly follow the network configuration of NCM~\cite{borisov2016neural} and adopt the gated recurrent unit (GRU)~\cite{cho2014properties} as the building block, which performs similarly to LSTM~\cite{hochreiter1997long} but is computationally cheaper. The process of generator $\pi_\theta (a|s)$ can be divided into following steps: 
\begin{itemize}[leftmargin=15pt]
    \item[(1)] A user starts the session by issuing query $q$ and the hidden state $\mathbf{h}_0$ is initialized with $q$, where the document $\mathbf{v}_d$, vertical type $\mathbf{v}_v$ and \emph{previous interaction} $\mathbf{v}_c$ are initialized with  $\mathbf{0}_d, \mathbf{0}_v, \mathbf{0}_c$. 
    \item[(2)] At rank $1$, the user examines the first document. The current hidden state $\mathbf{h}_1$ encodes the last state $\mathbf{h}_0$, document embedding for current document $\mathbf{v}_d$, its corresponding vertical type embedding $\mathbf{v}_v$, and the previous interaction $\mathbf{v}_c$ via a GRU unit. The current action $a_1$ with document $d_1$ is generated according to the action probability from the policy $\pi_\theta (a_{1} | s_{1}) = \operatorname{Softmax}\left(\operatorname{Linear}(\mathbf{h}_{1})\right)$; 
    \item[(3)] The previous interaction $\mathbf{v}_c$ is updated with the embedding of current action $a_1$, i.e., click or not click.
    \item[(4)]  For rank $t>1$, steps~(2) and (3) are repeated to select current action $a_t$ and update previous interaction $\mathbf{v}_c$.
\end{itemize}

The structure of the generator $\pi_\theta (a|s)$ is described as:
\begin{equation}
\begin{aligned}
&\mathbf{x}_{t}=\left\{\begin{array}{ll}\mathbf{v}_{q} \oplus \mathbf{0}_d \oplus \mathbf{0}_v \oplus \mathbf{0}_c & t=0 \\ \mathbf{v}_{q} \oplus \mathbf{v}_{d} \oplus \mathbf{v}_{v} \oplus \mathbf{0}_c & t=1 \\ 
\mathbf{v}_{q} \oplus \mathbf{v}_{d} \oplus \mathbf{v}_{v} \oplus \mathbf{v}_{c} & t=2, \ldots, T\end{array}\right.,\\ 
&\mathbf{h}_{t}=\operatorname{GRU}_{g}\left(\mathbf{h}_{t-1}, \mathbf{x}_{t}\right), \\ 
&\pi_\theta \left(a_{t} | s_{t}\right)=\operatorname{Softmax}\left(\operatorname{Linear}(\mathbf{h}_{t})\right),
\end{aligned}
\label{eq:generator}
\end{equation}
where $\oplus$ is the vector concatenation, $\mathbf{0}_*$ denotes a zero vector with size $l_*$ for the corresponding feature and $\mathbf{h}_{t}$ is the hidden representation of $s_t$. 
Note that in NCM the query embedding $\mathbf{v}_{q}$ is only used at step 0 while in AICM we encoded this information at each step to ensure it not forgotten during the propagation of RNN~\cite{graves2014neural}.


\subsection{Discriminator}
\label{sec:discriminator}
The discriminator $D_w(s,a)$ distinguishes the state-action pairs $(s,a)$ generated by the behavior policy $\pi_\theta(a|s)$ from those of expert policy. We also use GRU as the building block for $D_w(s,a)$. To be consistent with the generator, the initial state $\mathbf{h}_0^\prime$ of the discriminator is initialized with query embedding $\mathbf{v}_q$ at rank $0$. At rank $t\geq 1$, the GRU unit takes as input the query embedding $\mathbf{v}_q$, document embedding $\mathbf{v}_d$, vertical embedding $\mathbf{v}_v$, and \emph{current interaction} $\mathbf{v}_c$ with $d_t$ (recall that in Eq.~\eqref{eq:generator}, $\mathbf{v}_c$ is users' previous interaction with $d_{t-1}$), and outputs a hidden vector $\mathbf{h}_t^\prime$, which contains the information of both $s_t$ and $a_t$. The structure of the discriminator $D_w(s,a)$ is described as:
\begin{equation}
\begin{aligned}
&\mathbf{x}_{t}^{\prime}=\left\{\begin{array}{ll}\mathbf{v}_{q} \oplus \mathbf{0}_d \oplus \mathbf{0}_v \oplus \mathbf{0}_c & t=0 \\ \mathbf{v}_{q} \oplus \mathbf{v}_{d} \oplus \mathbf{v}_{v} \oplus \mathbf{v}_{c} & t=1, \ldots, T
\end{array}\right., \\ &\mathbf{h}_{t}^{\prime}=\operatorname{GRU}_{d}\left(\mathbf{h}_{t-1}^{\prime}, x_{t}^{\prime}\right), \\ 
&D_w\left(s_{t}, a_{t}\right)=\operatorname{Sigmoid}\left(\operatorname{Linear}\left(\mathbf{h}_{t}^{\prime}\right)\right).
\end{aligned}
\end{equation}
Note that the hidden state $\mathbf{h}_t^\prime$ encodes the information of both the state $s_t$ and action $a_t$ while $\mathbf{h}_t$ in Eq.~\eqref{eq:generator} only encodes the information of state $s_t$. 

\subsection{Adversarial Training}
The behavior policy $\pi_\theta(a|s)$, i.e., the generator, generates click sequences, while 
the discriminator $D_w(s, a)$ measures the difference between the generated click sequences and the ground-truth sequences. 
The generator and discriminator are updated according to the following procedures until convergence. 

Firstly, we sample trajectories $\tau_{\pi_\theta}$ from the behavior policy $\pi_\theta(a|s)$ and 
$\tau_{\pi_E}$ from expert demonstrations, then we update the discriminator parameters $w$ with the gradient  
\begin{equation}
    \hat{\mathbb{E}}_{\tau_{\pi_\theta}}[\nabla_{w}\log D_w(s,a) ]+\hat{\mathbb{E}}_{\tau_{\pi_E}}[\nabla_{w}\log(1-D_w(s,a))]\,.
\label{grad:d}
\end{equation}

Each time we obtain an updated discriminator $D_w(s,a)$, we take a gradient step using Proximal Policy Optimization (PPO)~\cite{schulman2017proximal} to update the generator, according to
\begin{equation}
    \hat{\mathbb{E}}_{\tau_{\pi_\theta}}[\nabla_{\theta}\log\pi_{\theta}(a|s)Q(s,a)]-\lambda\nabla_{\theta}H(\pi_{\theta})\,,
\label{grad:g}
\end{equation}
where
\begin{equation}
    Q(\Bar{s},\Bar{a})=\hat{\mathbb{E}}_{\tau_{\pi_\theta}}[\log D_{w}(s,a)|s_0=\Bar{s},a_0=\Bar{a}]\,.
\label{eq:q_function}
\end{equation}

The state-action value function defined in Eq.~\eqref{eq:q_function}, which controls the direction and scale of policy gradient, is built upon the reward provided by the discriminator $D_w(s,a)$. It works from two aspects. 
On one hand, it gives low reward when the next click of the generated click sequence differs from training data, which is similar to most state-of-the-art methods. On the other hand, it also gives low reward to the generated sequence where the prefix, i.e., previous generated clicks, is significantly different from training data,
which explicitly constrains the propagation of error. During the training, as the discriminator better distinguishes the generated sequences and the ground truth sequences, which makes the generator produce more realistic prefix, the exposure bias can be sufficiently alleviated. 

Overall, by alternately updating the discriminator and the generator according to Eq.~\eqref{grad:d} and Eq.~\eqref{grad:g}, we solve the optimization problem in Eq.~\eqref{eq:obj1}, which essentially minimizes the JS divergence between the occupancy measures under the behavior policy and the expert policy. Recall that, most existing methods minimize KL divergence, which tends to average on all the patterns (assuming that user behaviors are complex and contain multiple pattern modes) and thus fails to fit the complex user behaviors. Compared to KL divergence, minimizing JS divergence, which concentrates more on the major pattern~\cite{bishop2006pattern}, encourages each click generated by behavior policy to be ``real'' according to expert demonstrations instead of trying to correctly cover each pattern from the demonstrations. Such tendency ensures the good quality of generated clicks, especially in case of sparse data and large search space.

\section{Theoretical Analysis}
In this section, we analyze how the exposure bias is reduced in AICM theoretically. 

\begin{definition}
Define $R(s_t, a_t)$ as the user's immediate utility gain at step $t$.  $J(\pi) = \mathbb{E}_{s_0\sim p_0, a_t\sim\pi(\cdot|s_t), s_{t+1}\sim P(\cdot|s_{t}, a_{t})} [\sum_{t=0}^T \gamma^t R(s_t, a_t)]$ denotes the expected $T$ step utility under current click policy $\pi$.  
\label{def:1}
\end{definition}

Definition~\ref{def:1} defines the user's intrinsic reward function and the policy-level expected utility based on it. In click models, future states are influenced by current decisions.  
The discount factor $\gamma$ describes how much we should consider future states to make an optimal decision at each step. 
We assume that under the expert policy, the user's utility is maximized, so the ultimate goal for a click model is to minimize the following utility gap:
\begin{equation}
    \begin{aligned}
    &|J(\pi) - J(\pi_E)| \\
    = & 
    \big|\sum\nolimits_{t=0}^{T}  \gamma^t  \sum\nolimits_{s_t,a_t} (P_\pi(s_t, a_t) - P_{\pi_E}(s_t, a_t))R(s_t, a_t)\big|\,. \\
    \end{aligned}
\end{equation}

Firstly, we derive the utility gap for a click policy $\pi(a|s)$ based on behavior cloning. From Eq.~\eqref{eq:bcobj} we can derive the following theorem.


\begin{theorem}
    For a click policy $\pi(a|s)$ based on behavior cloning, the KL divergence between $\pi(a|s)$ and the expert policy for each state visited by the expert policy can be bounded by a constant $\epsilon_{bc}$, i.e., \\ $\max_s D_{KL} (\pi_E(\cdot|s), \pi(\cdot|s)) \leq \epsilon_{bc}.$ Assume that the reward function is bounded by an absolute value $R_{max}$, then the utility gap $|J(\pi) - J(\pi_E)|$ is bounded by
    \begin{equation}
    |J(\pi) - J(\pi_E)| \leq 2T(T+1) R_{max} \sqrt{\epsilon_{bc}}.
    \end{equation}
\label{thrm:bc}
\end{theorem}

The proof is in Appendix~\ref{proof:bc}. According to Theorem~\ref{thrm:bc}, the exposure bias problem exists for a BC-based click policy. To be specific, in training stage the policy $\pi(a|s)$ is learned assuming each previous click behavior is real while in testing stage the next click is generated based on previous predictions, which might have not been seen during training. Under such a condition the induced utility gap is quadratic w.r.t. the list length $T$.   

After that, we derive the utility gap for a click policy $\pi(a|s)$ based on GAIL. From Eq.~\eqref{eq:obj2} we can derive the following theorem.

\begin{theorem}
    For a click policy $\pi(a|s)$ based on GAIL, the JS divergence between the occupancy measure $\rho(s, a)$, i.e., the distribution of state-action pairs, under the click policy $\pi(a|s)$ and expert policy $\pi_{E}(a|s)$ can be bounded by $\epsilon_{ga}$, i.e., $
    D_{JS} (\rho_{\pi}, \rho_{\pi_E}) \leq \epsilon_{ga}.$ Assume that the reward function is bounded by an absolute value $R_{max}$, then the utility gap $|J(\pi) - J(\pi_E)|$ is bounded by
    \begin{equation}
        |J(\pi) - J(\pi_E)| \leq 2 \sqrt{2} R_{max} (T+1) \sqrt{\epsilon_{ga} }.
    \end{equation}
\label{thrm:gail}
\end{theorem}

The proof is in Appendix~\ref{proof:gail}. According to Theorem~\ref{thrm:gail}, the utility discrepancy induced by AICM is linear to list length $T$. AICM generates each click based on previous predictions and evaluates the quality of the whole generated sequence instead of one-step click. 
In such a dynamic training, we alleviate the exposure bias as in BC-based methods and reduce the utility gap significantly from $\mathcal{O}(T^2)$ to $\mathcal{O}(T)$.
 

%% file: text/exp.tex
\section{Experiment}

In this section, we conduct extensive experiments\footnote{The experiment code with running instructions is available at \url{https://github.com/xydaisjtu/AICM}.} to answer the following questions:

\begin{itemize}
\item[\textbf{RQ1}]
 How does AICM perform in click prediction and relevance estimation compared with the existing click models?
\item[\textbf{RQ2}]
 Does AICM perform better than the existing click models in recovering the distribution of real data?
\item[\textbf{RQ3}]
 How does AICM perform when the document lists are provided in a bad order?
 \item[\textbf{RQ4}]
 What are the influences of different model configurations?
\end{itemize}

\subsection{Experimental Setup}

\subsubsection{Dataset}

Following CACM~\cite{chen2020context}, we use TianGong-ST dataset\footnote{\url{http://www.thuir.cn/tiangong-st/}}, which is an open search log dataset released by Chinese commercial search engine \textit{Sougou.com} in 2019. In TianGong-ST~\cite{chen2019tian}, there are 147,155 sessions, 340,596 unique queries and 309,287 unique documents. The dataset is divided into training, validation, and test sets with proportion 8:1:1. There are also 2,000 query-sessions with human-annotated relevance labels to facilitate relevance estimation. The statistics of this dataset are shown in Table~\ref{tab:Data segmentation}.

\begin{table}[h]
    \centering
    \caption{Basic statistics of TianGong-ST dataset.}
    \label{tab:Data segmentation}
	\vspace{-9pt}
	\scalebox{1.0}{
	\renewcommand\arraystretch{1.05}
    \begin{tabular}{c c c c}
    \toprule
     & training & validation & test \\
    \midrule
    \# sessions & $117,431$ & $13,154$ & $16,570$ \\
    \# unique queries & $35,903$ & $9,373$ & $11,391$ \\
    avg. session length & $2.4099$ & $2.4012$ & $2.4986$ \\
    \bottomrule
    \end{tabular}
    }
\vspace{-10pt}
\end{table}

\subsubsection{Baselines}

The existing click models can be categorized into two classes: PGM-based and NN-based methods. CCM~\cite{Guo2009CCM}, DCM~\cite{Guo2009DCM}, DBN~\cite{Chapelle2009DBN}, SDBN~\cite{chuklin2015click}, PBM~\cite{craswell2008experimental} and UBM~\cite{dupret2008user} are considered as representative PGM-based click models, of which open-source implementations are available\footnote{\url{https://github.com/markovi/PyClick}}. For NN-based click models, we consider NCM~\cite{borisov2016neural} and CACM~\cite{chen2020context} for experimental comparison.

\subsubsection{Evaluation Metrics}

We use three traditional metrics for click prediction and relevance estimation tasks. In addition, we propose two metrics (Reverse PPL and Forward PPL) to evaluate the generalization and data distributional coverage of click models. More details are described in Section~\ref{sec:Distributional Coverage}. 

For click prediction task, we report the log-likelihood (LL) and perplexity (PPL)~\cite{dupret2008user} of each model. The definitions of the log-likelihood and click perplexity at the rank $r$ are as follows:
\begin{align}
    &LL = \frac{1}{MN}\sum_{i=1}^{N}\sum_{t=1}^{M}C_{i,t}\log \mathcal{P}_{i,t}+(1-C_{i,t})\log(1-\mathcal{P}_{i,t}),\\
    &PPL@t=2^{-\frac{1}{N}\sum_{i=1}^{N}C_{i,t}\log \mathcal{P}_{i,t}+(1-C_{i,t})\log(1-\mathcal{P}_{i,t})},
\end{align}
where the subscript $t$ 
is the rank position in a result list, $N$ is the total number of queries, and $M$ is the number of results in a query. $C_{i,t}$ and $\mathcal{P}_{i,t}$ denote the real click signal and the predicted click probability of the $t$-th result in the $i$-th query. The total perplexity performance is calculated by averaging perplexities over all the positions. Lower values of perplexity and higher values of log-likelihood correspond to better click prediction performance. 

For relevance estimation task, we use click models to rank the document list and calculate the mean Normalized Discounted Cumulative Gain (NDCG)~\cite{NDCG} according to the human labels. We report NDCG scores at truncation level 1, 3, 5 and 10.

\subsubsection{Implementation Details}
\label{implement}
We train AICM with a mini-batch size of 128 by using the Adam optimizer. The embedding size and hidden size of GRU are both 64. The initial learning rate for the generator and discriminator are $5\times 10^{-4}$ and $1\times 10^{-3}$ with a decay rate of $5\times 10^{-1}$. To avoid overfitting, we set the coefficient of L2 norm and dropout rate to $1\times 10^{-5}$ and $5\times 10^{-1}$. At the beginning of the training, we use the maximum likelihood estimation (MLE) to pre-train the generator $\pi_\theta$ and discriminator $D_w$ on training set with the initial learning rate of $1\times 10^{-3}$. Finally, we adopt the model at the iteration with the lowest validation PPL for evaluation in the test set. To ensure fair comparison, we also fine-tune all the baseline models to achieve their best performance.

\subsection{Performance on Traditional Metrics (RQ1)}
The results for the click prediction task and the relevance estimation task are presented in Table~\ref{tab:Overall performance}, from which we can obtain the following observations.

\begin{table}[t]
	\centering
	\caption{Click model performance on traditional metrics. The best results are given in bold. $*$ and $**$ indicate statistically significant improvement (measured by Wilcoxon signed-rank test with p-value$<$ 0.01 and p-value$<$0.001) over all baselines.}
	\label{tab:Overall performance}
	\resizebox{\columnwidth}{!}{
	\renewcommand\arraystretch{1.11}
	\begin{tabular}{c|c c|c c c c }
		\hline
		\multicolumn{1}{c|}{\multirow{2}{*}{Model}} & \multicolumn{2}{c|}{Click Prediction} & \multicolumn{4}{c}{Relevance Estimation} \\ 
		\cline{2-7} 
		\multicolumn{1}{c|}{} & LL & PPL & NDCG@1 & NDCG@3 & NDCG@5 & NDCG@10 \\
		\cline{1-7}
		\multicolumn{1}{c|}{CCM} & -0.2224  & 1.2034 & 0.6702 & 0.6941 & 0.7229 & 0.8477 \\
		\multicolumn{1}{c|}{DCM} & -0.2302 & 1.1994 & 0.6807 & 0.6824 & 0.7161 & 0.8452 \\
		\multicolumn{1}{c|}{DBN} & -0.2218 & 1.2103 & 0.6711 & 0.6958 & 0.7241 & 0.8471 \\
		\multicolumn{1}{c|}{SDBN} & -0.2328 & 1.2116 & 0.6868 & 0.6846 & 0.7177 & 0.8455 \\
		\multicolumn{1}{c|}{PBM} & -0.1483 & 1.1894 & 0.6481 & 0.6419 & 0.6726 & 0.8235 \\
		\multicolumn{1}{c|}{UBM} & -0.1494 & 1.1896 & 0.6435 & 0.6381 & 0.6681 & 0.8223 \\
		\hline
 		\multicolumn{1}{c|}{NCM} & -0.1443 & 1.1855 & 0.7003 & 0.7041 & 0.7351 & 0.8608 \\
 		\multicolumn{1}{c|}{CACM} & -0.1426 & 1.1832 & 0.7347 & 0.7153 & 0.7403 & 0.8662 \\
 		\multicolumn{1}{c|}{AICM} & $\textbf{-0.1385}^{**}$ & $\textbf{1.1747}^{**}$ &\textbf{0.7348} & $\textbf{0.7167}^*$ & $\textbf{0.7439}^*$ & $\textbf{0.8667}^*$ \\
 		\hline
	\end{tabular}
	}
\end{table}

\begin{itemize}[leftmargin=15pt]
    \item[(1)] All NN-based models significantly outperform PGM-based models in the click prediction and the relevance estimation tasks. NN-based models learn the distributed representations of queries and documents, therefore they can better capture the user behavior patterns.
    \item[(2)] CACM performs the best among all the baseline models, followed by NCM and PGM-based click models. CACM can better capture the user behavior patterns by taking session-level information into account, which is consistent with the results reported in~\cite{chen2020context}.
    \item[(3)] For click prediction, AICM significantly outperforms all the baseline models. As AICM shares the same policy structure with NCM, such improvement validates the effectiveness of applying GAIL framework on click models for click prediction. Adversarial training enables AICM to better capture the user  behavior patterns, instead of only fitting the click logs via MLE.
    \item[(4)] For relevance estimation, AICM performs better than 
    the best baseline model (i.e., CACM) at all truncation levels in terms of NDCG. 
    The baseline CACM models examination prediction and relevance estimation separately, and uses the extra session-level information. On the contrary, without complex model structure and session-level side information, our AICM can still achieve comparatively better performance with CACM in relevance estimation task, which demonstrates the effectiveness of our proposed GAIL framework for click models. 
\end{itemize}

\subsection{Distributional Coverage (RQ2)}
\label{sec:Distributional Coverage}

\subsubsection{Metrics for Distributional Coverage}

Regarding the click model as a type of generative model for click signal generations, click models are to approximate the true data distribution that underlies the click log data and generate click samples of high fidelity. Traditional tasks and the corresponding metrics (e.g., click prediction task and LL, PPL) are not very suitable, because they view the click model as a predictive model and only deal with one-step click probabilities $\mathcal{P}_{i,t}$ conditioned on true previous clicks.
We need a task that views the click model as a generative model and measures the quality of generated samples, i.e., the whole click sequences based on its own predictions. 
Therefore, we propose a novel task for click models, called \textbf{distributional coverage}, in which we aim to measure the similarity between the true data distribution and the data distribution learned by the click model. Greater similarity between the true data distribution and the learned data distribution implies higher fidelity and better distributional coverage of the click model.

Since it is not possible to obtain the true data distribution, we cannot measure the similarity directly. A common quantitative measure to test
the fidelity and distributional coverage of generative models is to evaluate the generated samples via a strong surrogate model. Following~\cite{Zhao2018WAE}, we propose two novel metrics, \textbf{Reverse PPL} and \textbf{Forward PPL}. 
The surrogate model is the intermediary to evaluate the similarity between the generated samples and the real data samples.  
Reverse PPL is the PPL of a surrogate model that is trained on generated samples and evaluated on held-out real data. Forward PPL is the PPL of a surrogate model that is trained on held-out real data and evaluated on generated samples. 

\begin{figure}[t]
	\centering
	\includegraphics[width=0.45\textwidth]{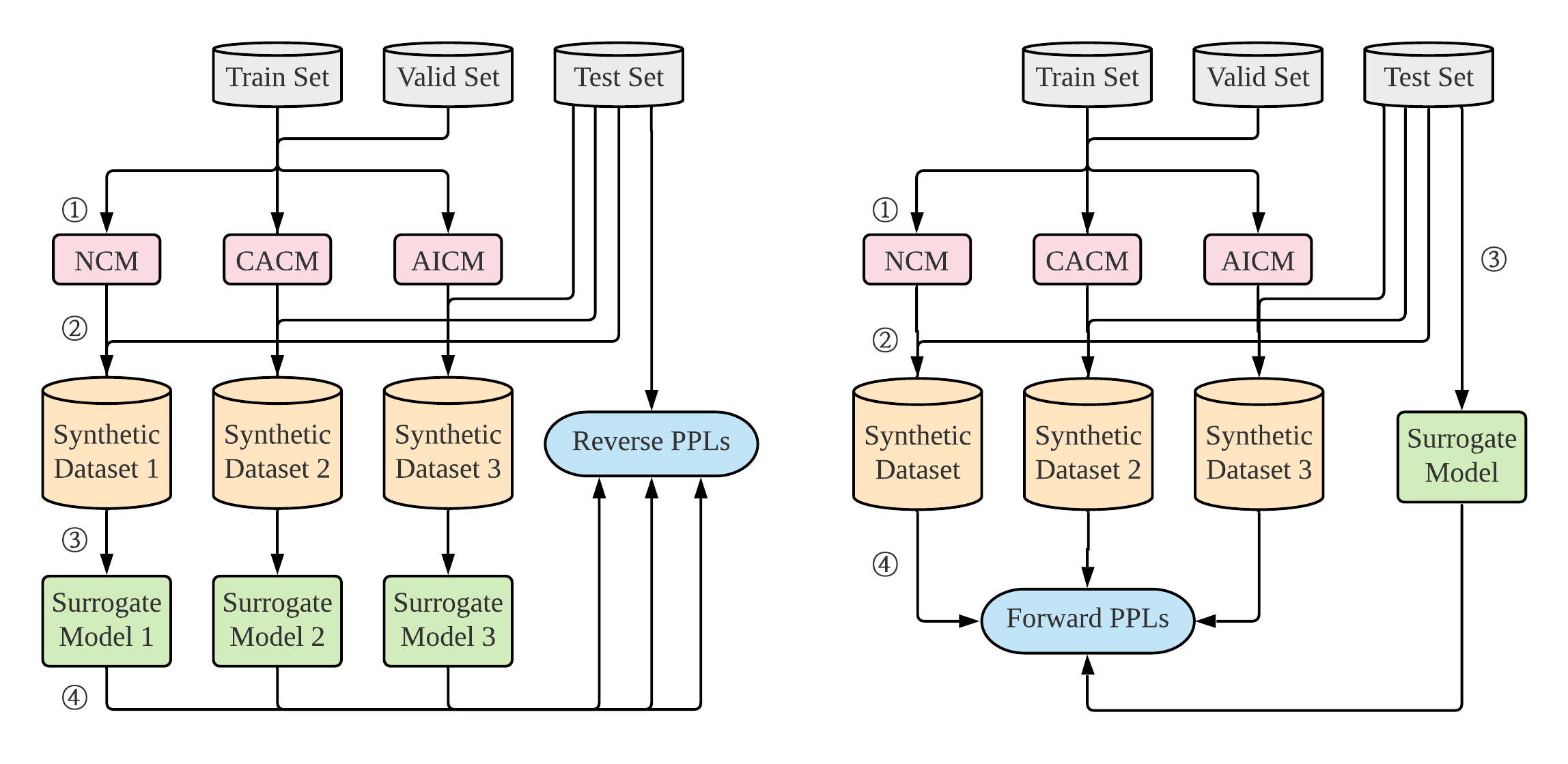}
	\caption{Computation Flow charts of Reverse PPL (Left) and Forward PPL (Right).}
	\label{fig:Reverse Foward PPL Flow Chart}
\end{figure}

The computation flow charts of Reverse PPL and Forward PPL are shown in Figure~\ref{fig:Reverse Foward PPL Flow Chart}. After training target click models based on training and validation sets, we use target click models to generate click signals based on queries and corresponding document lists in the test set, while document lists are kept in the original order (i.e., no permutation). Click signals are generated by sampling from a Bernoulli distribution that takes 1 with probability $\mathcal{P}_{i,t}$ and 0 with probability $1-\mathcal{P}_{i,t}$.

To generate a synthetic dataset of a similar size to the training set, click signals are independently sampled 7 times for each query in the test set, resulting in 289,835 queries. For each click model (e.g., NCM, CACM, AICM), a synthetic dataset is generated following the above process. To evaluate the fidelity and distributional coverage of different click models, we compute Reverse PPL and Forward PPL of individual synthetic datasets.
Lower value of Reverse/Forward PPL indicates better performance in distributional coverage task.

While traditional PPL metric in click prediction task only considers the click model as a predictive model, Reverse/Forward PPL in distributional coverage task view the click model as a generative model and directly measure the data distribution similarity by taking generated click sequences into account. Therefore, Reverse/Forward PPL are more suitable in real-world application scenarios where click models aim to build a simulation environment and provide simulated click signals.

\subsubsection{Performance for Distributional Coverage}

In our experiments, we measure the Reverse/Forward PPL of AICM, CACM, NCM and UBM. To conduct an adequate experiment, we use UBM as the PGM-based surrogate model and NCM as the NN-based surrogate model. For a fair comparison, surrogate models used in both Reverse PPL and Forward PPL for different click models are all of the same model size and are trained with the same training epochs. In addition, we also test Reverse/Forward PPL of the real data, i.e., the PPL of the surrogate model trained on held-out real data and evaluated on the same held-out real data, which are supposed to provide the best values for these two metrics. Results are presented in Table~\ref{tab:RQ2 Reverse Forward PPL}, from which we can obtain the following observations. 

\begin{itemize}[leftmargin=15pt]
    \item[(1)] As a PGM-based model, UBM achieves worse performance compared to NN-based methods in terms of Reverse PPL and Forward PPL for both two surrogate models, though its traditional PPL metric is very close to the best baseline model CACM in Table~\ref{tab:Overall performance}. This observation suggests that, the distributed vector representations is better than the traditional binary random variables representation in recovering the underlying distribution of click log data.
    \item[(2)] CACM fails to defeat NCM in these two metrics, though it shows a significant improvement in the traditional PPL metric. Observation (1) and (2) show that Reverse/Forward PPL for distributional coverage task have different tendences from traditional metrics (i.e., LL and PPL) for click prediction task. These two tasks evaluate different aspects of click models. Click prediction task considers one-step conditioned click probability, and distributional coverage measures the distributional discrepancy after the whole sequence is generated.
    \item[(3)] AICM outperforms all the baselines by a statistically significant margin ($p$-value < 0.001) in terms of both Reverse PPL and Forward PPL, with different surrogate models. This indicates that AICM can better recover the real data distribution of the click logs, which is to say, AICM is able to better capture the pattern of user behaviors in the real data.
    \item[(4)] An interesting observation we find in Table~\ref{tab:RQ2 Reverse Forward PPL} is that Forward PPL of AICM even outperforms that of the real data. On one hand, this observation indicates that AICM learns relatively simpler data pattern compared to the real data pattern, which can be regarded as a denoising process (i.e., outliers are removed). On the other hand, the ``proper'' performance of AICM for Reverse PPL (i.e., better than all the baselines and worse than the real data) shows that the data pattern learned by AICM is not too simple to fall in mode collapse.
\end{itemize}

\begin{table}[t]
    \centering
    \caption{Reverse/Forward PPL of Surrogate UBM/NCM models based on different synthetic datasets generated from target click models (e.g., UBM/NCM/CACM/AICM). Differences in Reverse/Forward PPL between any pair of the click models are statistically significant ($p$-value < 0.001).}
	\label{tab:RQ2 Reverse Forward PPL}
	\scalebox{0.75}{
	\renewcommand\arraystretch{1.05}
    \begin{tabular}{c|c|c}
    \hline
     & Surrogate UBM & Surrogate NCM \\
    \hline
        \begin{tabular}{c}
            Data \\
            \hline
            Real data \\
            UBM samples \\
            NCM samples \\
            CACM samples \\
            AICM samples \\
        \end{tabular} &
        \begin{tabular}{c c}
            Reverse PPL & Forward PPL \\
            \hline
            1.1412 & 1.1412 \\
            1.4249 & 3.3833 \\
            1.1831 & 1.2072 \\
            1.1854 & 1.2615 \\
            \textbf{1.1747} & \textbf{1.1383} \\
        \end{tabular} &
        \begin{tabular}{c c}
            Reverse PPL & Forward PPL \\
            \hline
            1.1453 & 1.1453 \\
            1.4231 & 2.9435 \\
            1.1848 & 1.2021 \\
            1.1812 & 1.2565 \\
            \textbf{1.1745} & \textbf{1.1324} \\
        \end{tabular} \\
    \hline
    \end{tabular}
    }
    \vspace{-10pt}
\end{table}

\subsubsection{Visualization for Distributional Coverage}

\begin{figure}[h]
	\centering
	\includegraphics[width=0.235\textwidth]{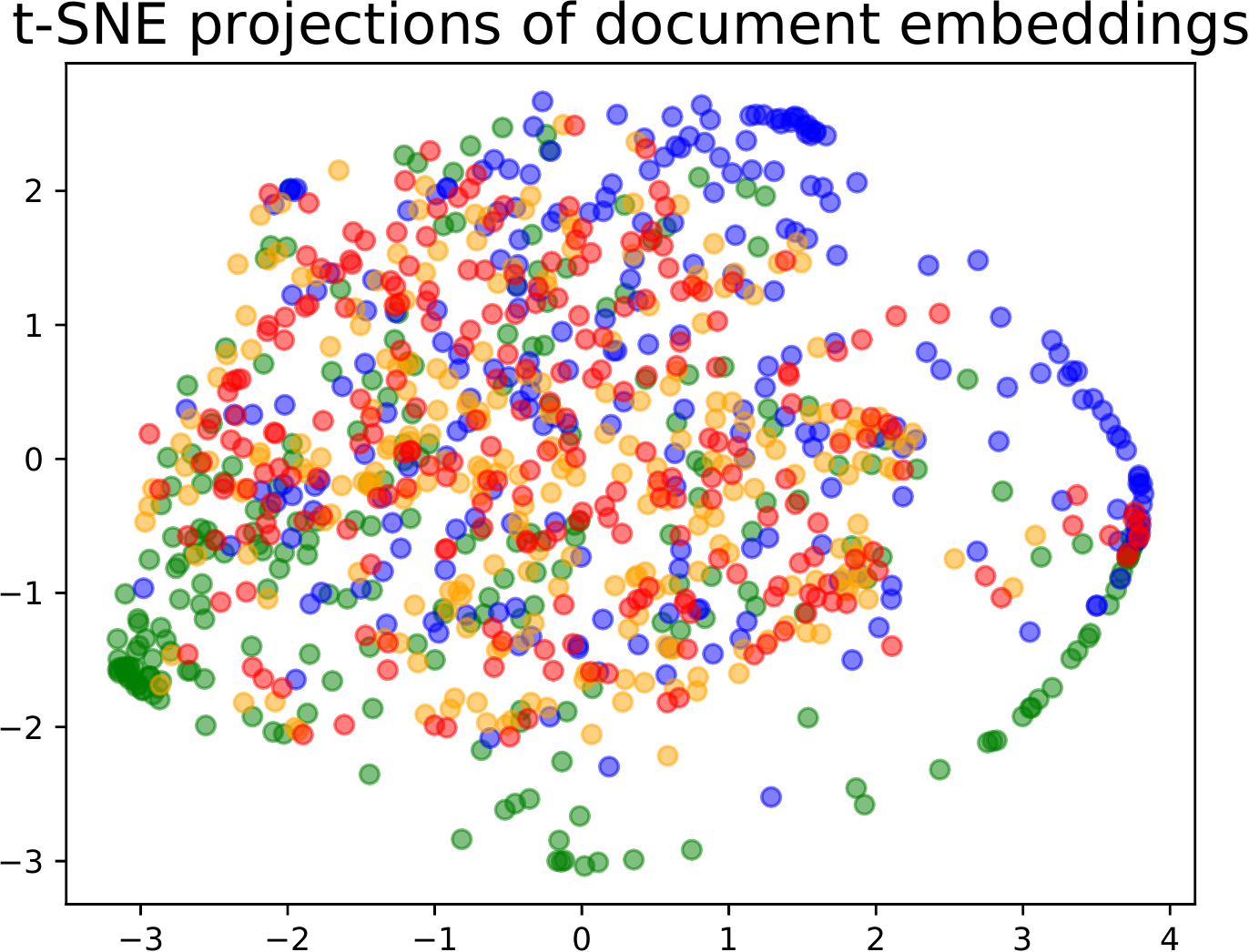}
	\includegraphics[width=0.227\textwidth]{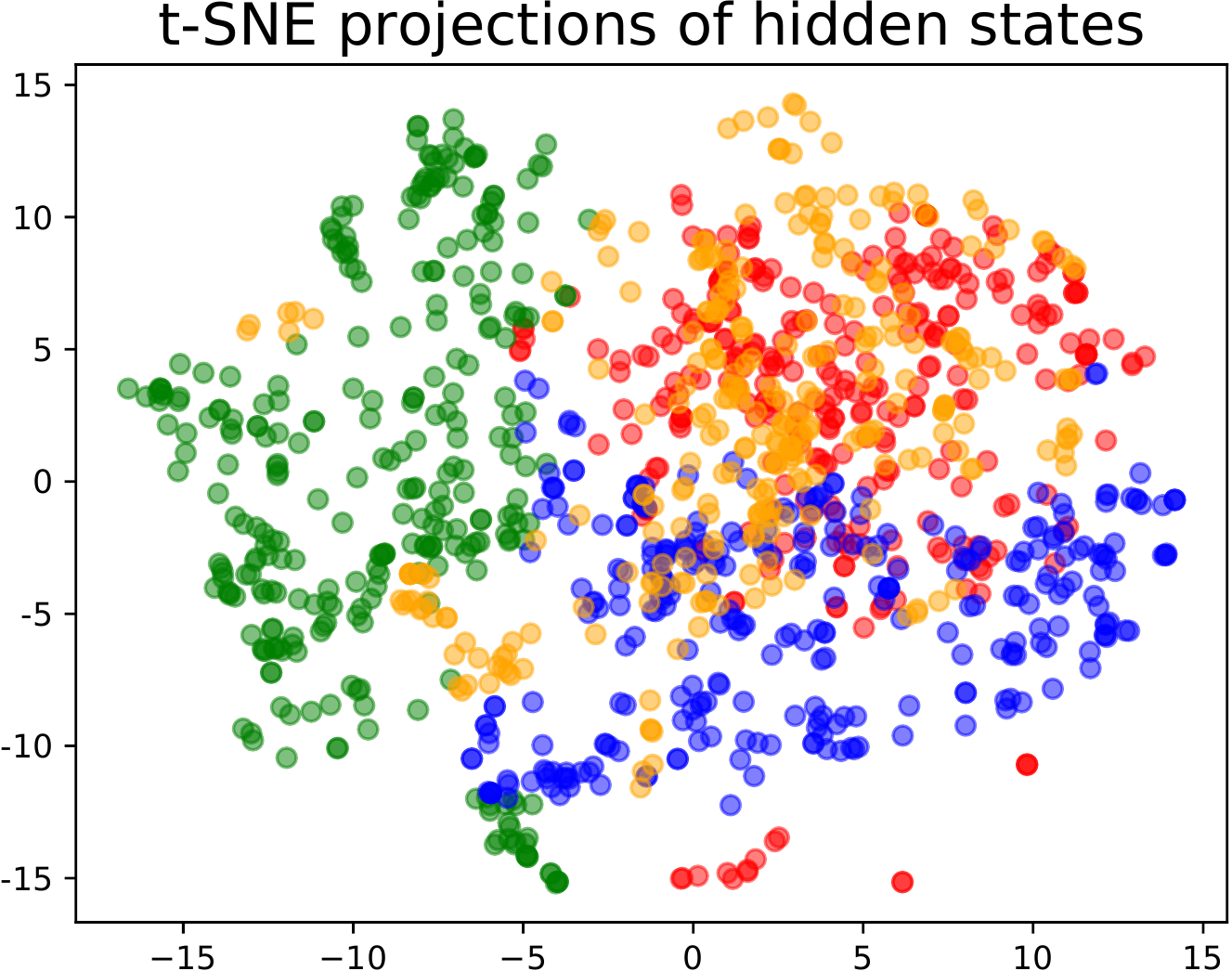}
	\caption{Two-dimensional t-SNE projections of document embeddings and hidden states learned by surrogate NCM model from different synthetic datasets. Colors correspond to synthetic datasets. Green: NCM, Blue: CACM, Orange: AICM, Red: real data. (Best viewed in color.)}
	\label{fig:t-SNE visualization}
\end{figure}

Furthermore, in Figure~\ref{fig:t-SNE visualization}, we visualize the t-SNE projections of the document embeddings and GRU hidden states learned by the surrogate NCM from synthetic datasets generated by different click models. The results on UBM synthetic dataset are not visualized because its Reverse/Forward PPL are significantly worse than the others. Note that we do not distinguish hidden states at different ranks with different colors. We can observe that both projections of document embeddings and GRU hidden states based on AICM synthetic dataset are closer to the real data compared to NCM and CACM. The projections of NCM and CACM perform similarly. These observations are consistent to the results of Reverse PPL in Table~\ref{tab:RQ2 Reverse Forward PPL}, which again validates the ability of AICM to capture the underlying distribution of user behaviors and generate click samples of high fidelity.

\subsection{Performance in Bad Cases (RQ3)}
Click models are trained and tested on real-world click logs, the document lists of which come from a well-trained ranking policy. 
However, when the click model is used as a simulation environment for a ranking policy, we cannot make the assumption that the ranking policy is always well-trained. Therefore, we consider whether the click model can provide stable performance in such bad cases where the document lists are not reasonably ranked. To be specific, we shuffle the original document lists which are well ranked and generate clicks based on such new lists. 

The traditional metrics (e.g., LL and PPL) are not suitable to evaluate the new generated clicks, because we do not have ground truth click signals on shuffled document lists. Whereas our proposed Reverse/Forward PPL in Section~\ref{sec:Distributional Coverage} are competent.

\begin{figure}[h]
    \centering
    \subfigure{
        \includegraphics[width=0.23\textwidth]{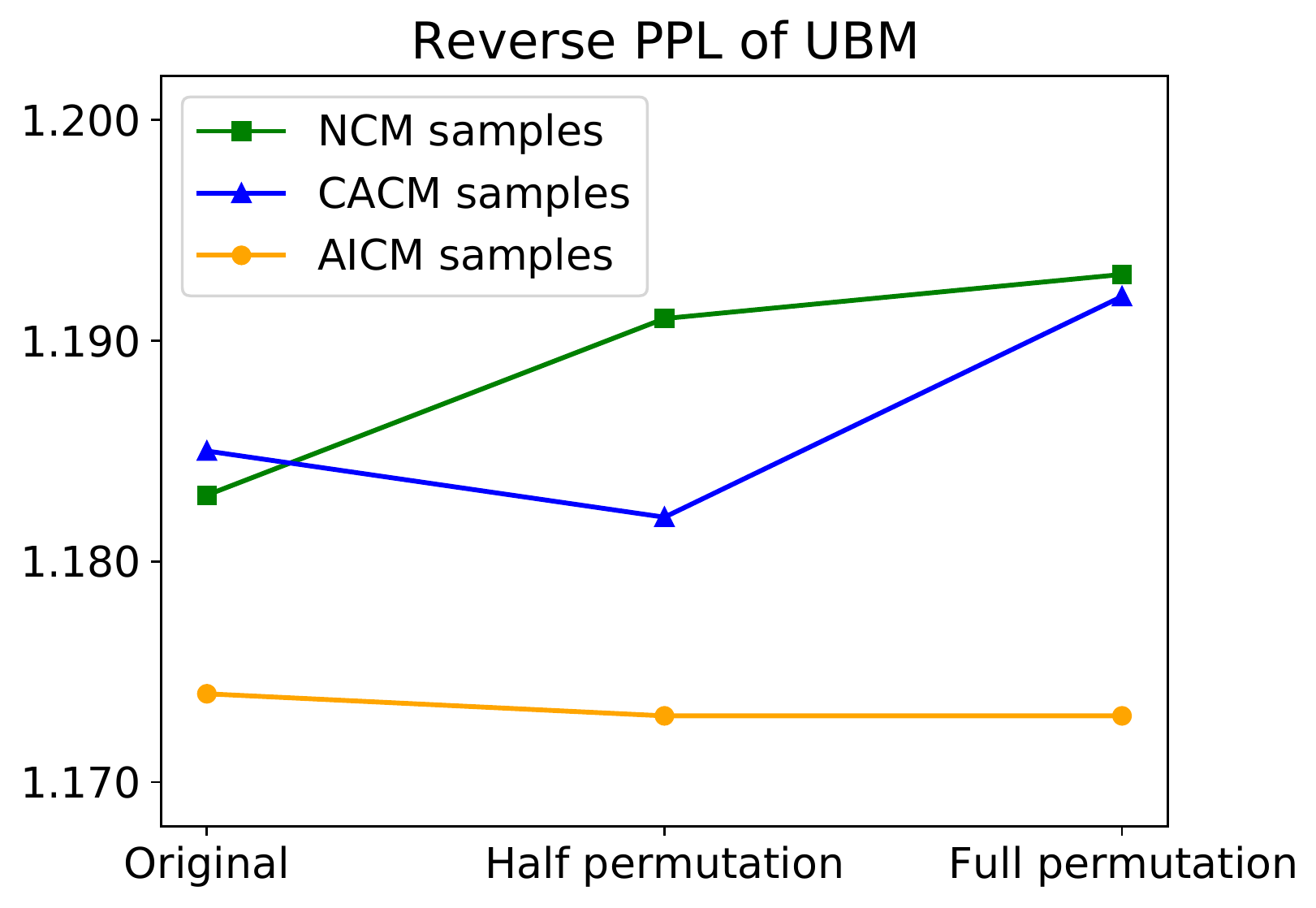}
        \includegraphics[width=0.23\textwidth]{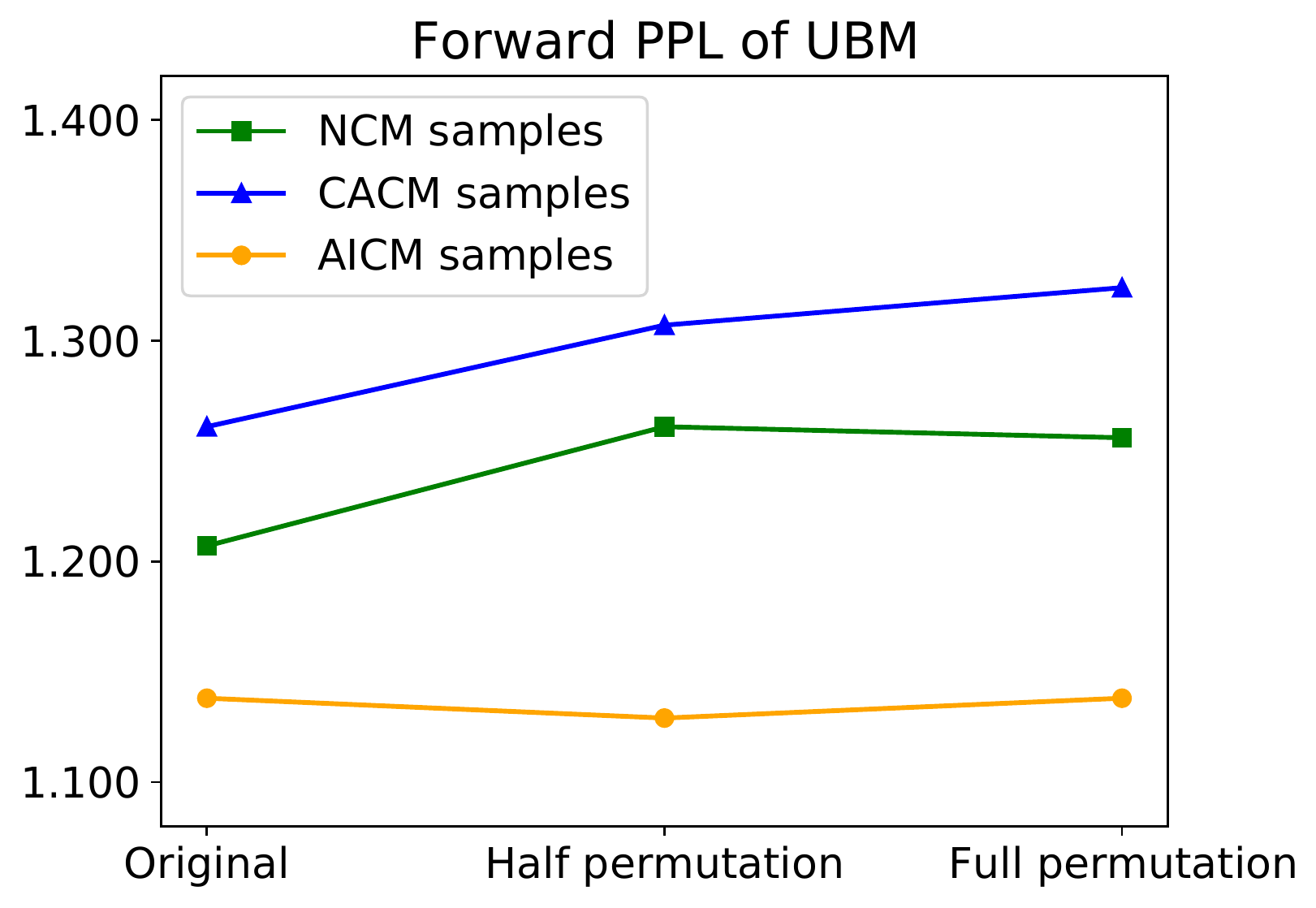}
    }
    \subfigure{
        \includegraphics[width=0.23\textwidth]{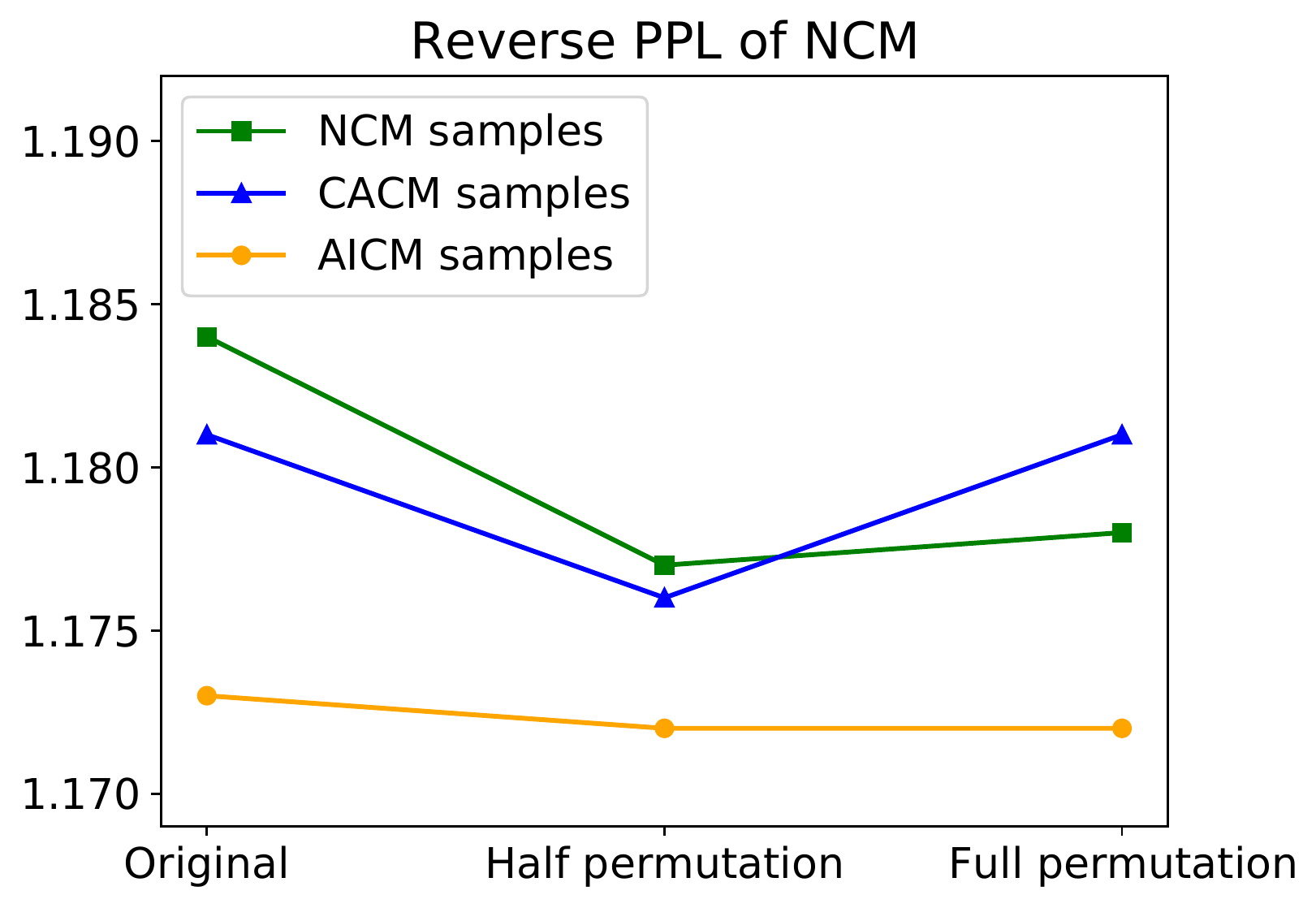}
        \includegraphics[width=0.23\textwidth]{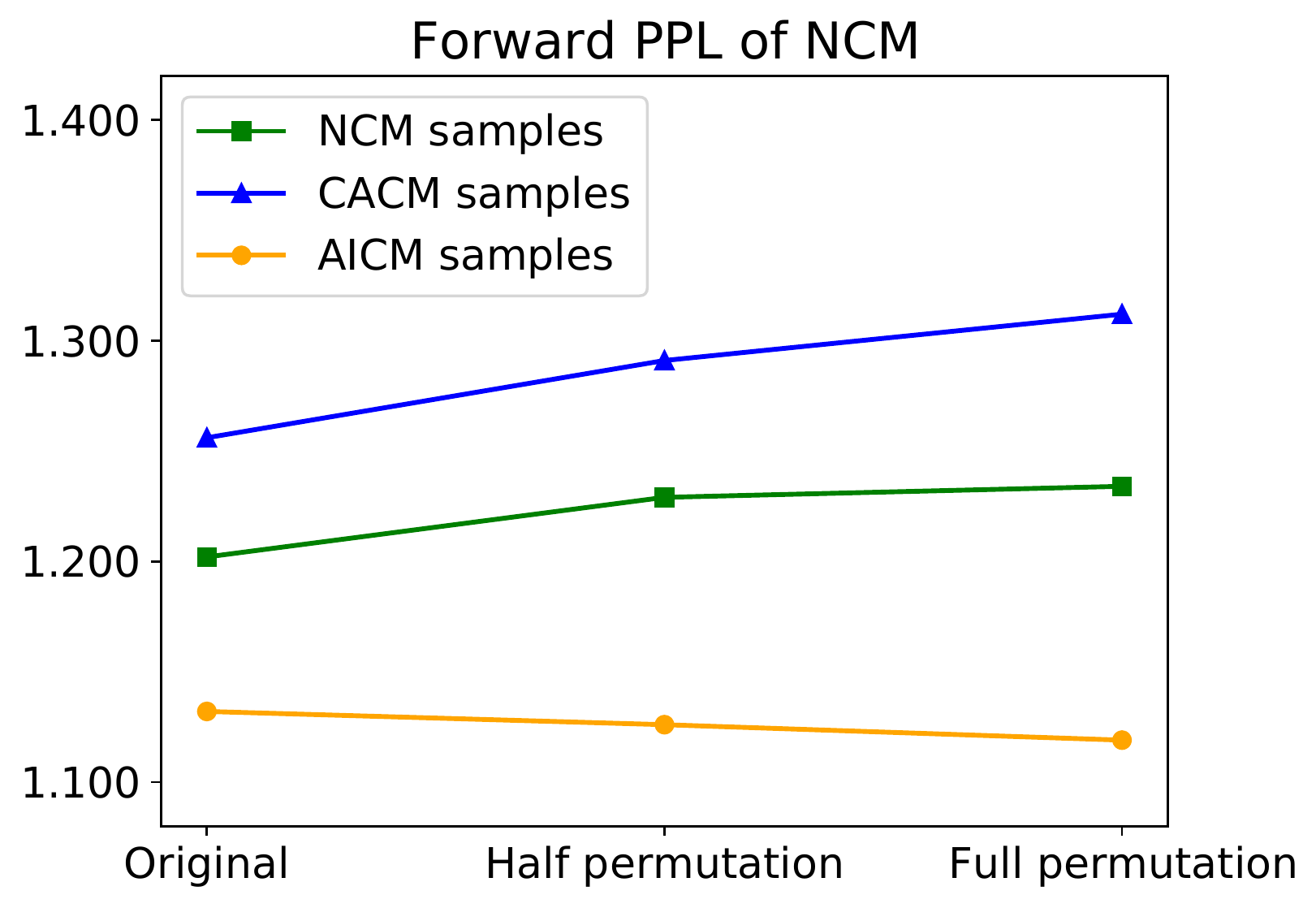}
    }
    \caption{Reverse PPL and Forward PPL of Surrogate UBM/NCM models based on different synthetic datasets at different permutation levels.}
    \label{fig:bad cases}
\end{figure}

Similar to that in Section~\ref{sec:Distributional Coverage}, we generate different synthetic datasets using different target click models, where the input document lists are permuted from the original test set. We permute the original document lists in two different ways: half permutation and full permutation. In half permutation, we separately shuffle the first half (i.e., rank 1 to 5) and the second half of the list, ensuring that the position of a document do not change dramatically (e.g., changing from rank 10 to rank 1). In full permutation, we shuffle the whole list, so that a striking position change is allowed. After generating synthetic datasets, we train surrogate models to measure Reverse/Forward PPL. The results are displayed in Figure~\ref{fig:bad cases}, from which we can obtain the following observations: 

\begin{itemize}[leftmargin=15pt]
    \item[(1)] Compared to NCM and CACM, AICM achieves the best and the most stable performance, no matter when the inputs are not permuted, half permuted and fully permuted.
    This indicates that, no matter whether the input lists are permuted, the data distribution of the synthetic dataset generated by AICM is consistently closest to the the real data distribution. This demonstrates that AICM is able to capture and simulate user behaviors even when it faces such bad cases where input lists are not well ranked.
    \item[(2)] We can sometimes observe performance improvement (i.e., the decrease of Reverse/Forward PPL) when input lists are permuted. Such phenomenon contradicts with our initial intuition that the performance of a click model should decrease if input lists are not well ranked, i.e., are permuted. The reason for this phenomenon differs in AICM and baselines.
    \begin{itemize}
        \item NN-based baseline models use MLE methods to cover the average pattern underlying the training set, which may sacrifice the generalization. If permuted lists happen to compensate for the pattern shrinkage caused by MLE methods, then it may lead to performance improvement in Reverse PPL, since Reverse PPL favors a diversified pattern. That is why in NN-based baseline methods, Reverse PPL sometimes decreases but Forward PPL does not. 
        \item AICM optimizes JS divergence, which is equivalently a combination of reverse KL divergence and forward KL divergence. Reverse KL divergence targets on the major pattern and forward KL divergence targets on the average pattern~\cite{bishop2006pattern}. Optimizing them together allows AICM to capture the underlying distribution properly, leading to a quite stable or even better performance on Reverse and Forward PPL when list permutation happens.
    \end{itemize}
\end{itemize}

\subsection{Ablation Study (RQ4)}

\subsubsection{Ablation Study on Pre-training Strategy}
\label{sec:ablation/pretrain}
\begin{figure}[t]
    \centering
    \includegraphics[width=0.23\textwidth]{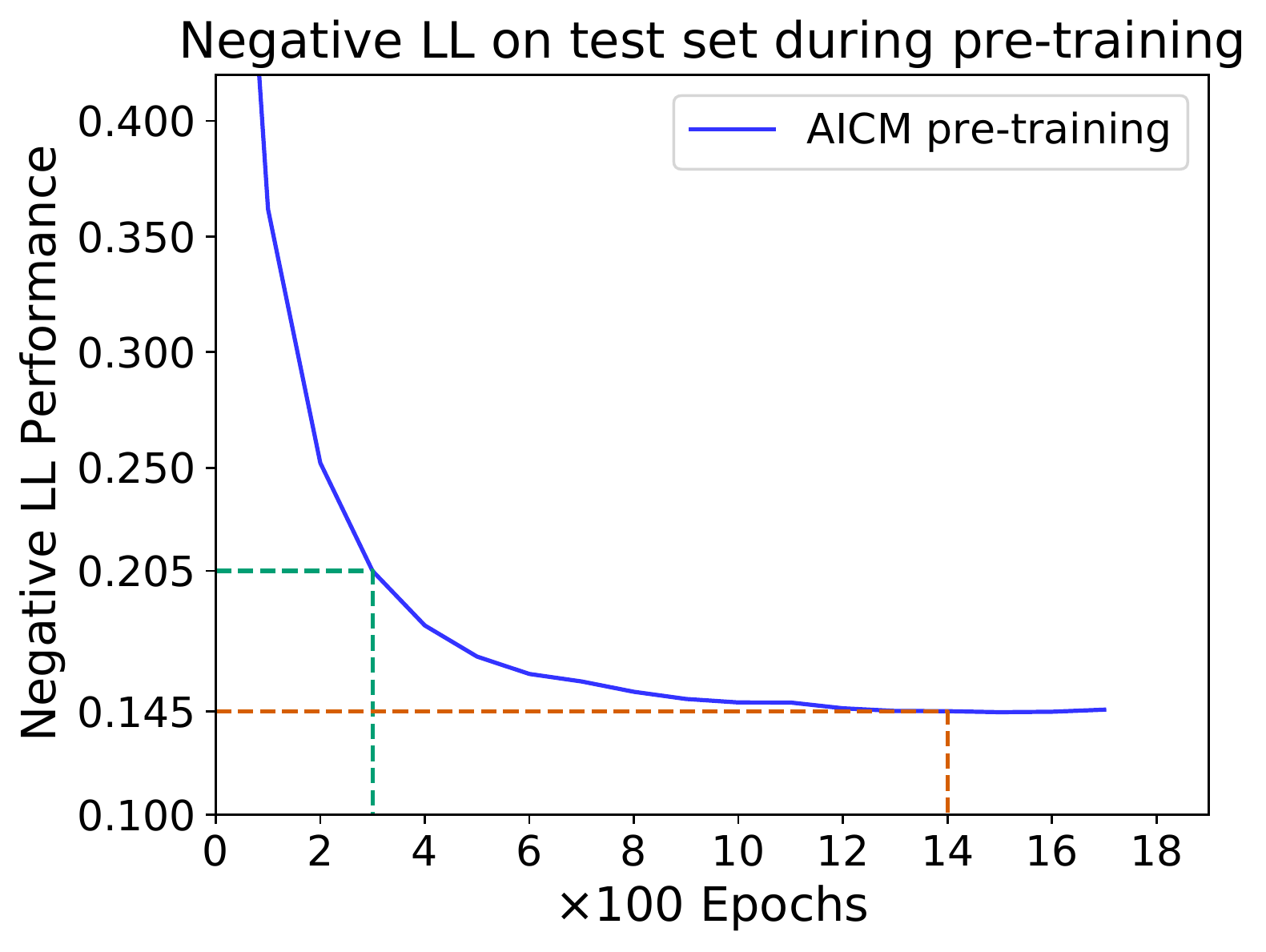}
    \includegraphics[width=0.2262\textwidth]{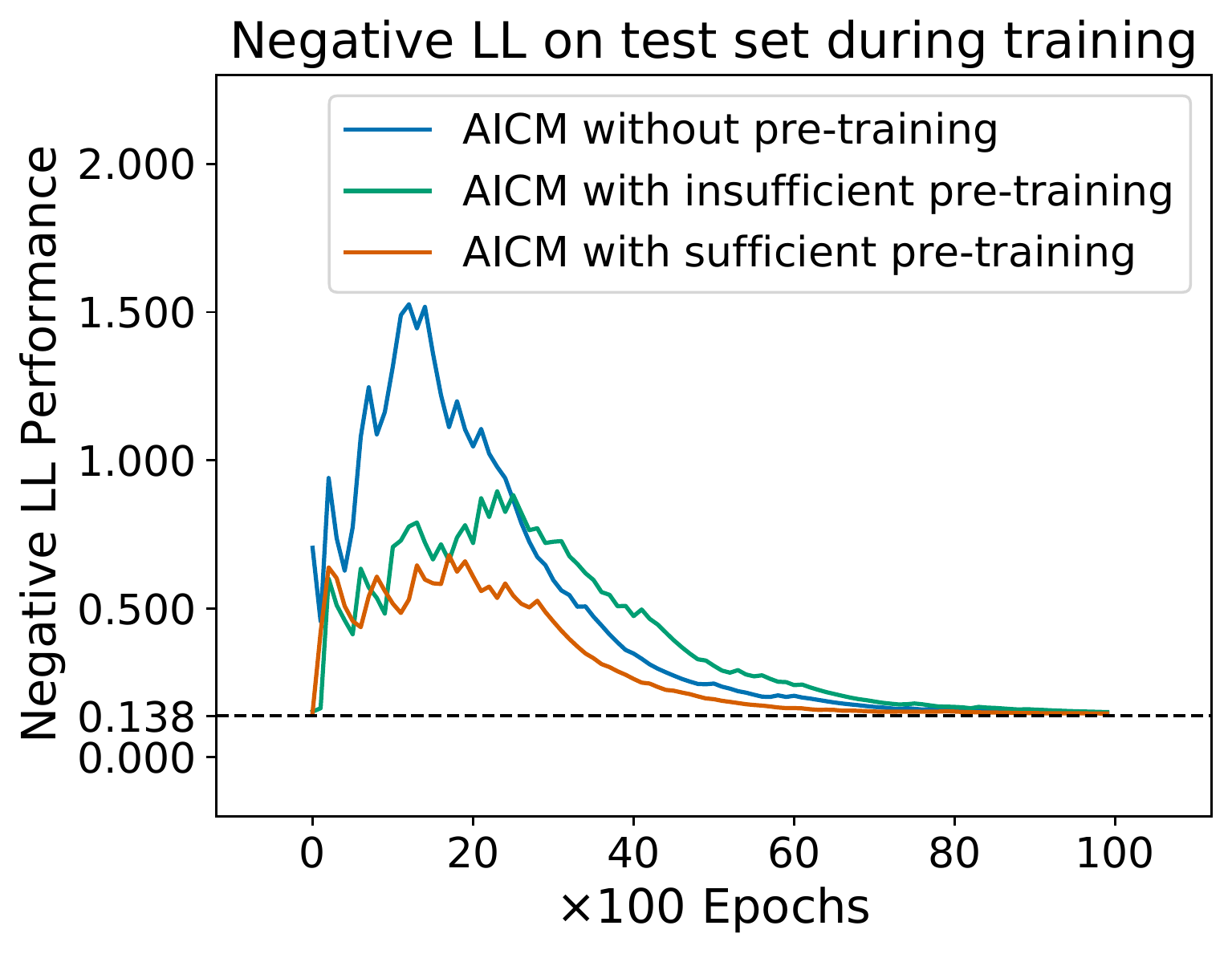}
    \caption{Left: Negative LL performance on test set during pre-training. The dashed lines represent the starting point of training AICM. Right: Negative LL performance on test set with different pre-training epochs before training. The dashed line represents the average negative LL performance over three training strategies. (Best viewed in color.)}
    \label{fig:ablation on pretrain}
\end{figure}

A sufficient pre-training is necessary to apply adversarial training to sequence generative models~\cite{yu2017seqgan}. In our experiments above, we also adopt the pre-training strategy to stabilize the adversarial training process. In this section, we conduct experiments to investigate the performance of AICM when the supervised pre-training is insufficient. The results are shown in Figure~\ref{fig:ablation on pretrain}. Only training curves of negative LL performance are displayed since all metrics (i.e., LL, PPL and NDCG) show a similar trend. We observe that the pre-training strategy does not influence the final convergence of AICM, but only impacts the range of performance fluctuation during training. 

The discriminator provides reward guidance when training the generator. 
If no pre-training strategy is applied or AICM is insufficiently pre-trained, the generator will act almost randomly at the beginning of the training, and the discriminator can identify the generated click sequences to be unreal with high confidence. 
This leads to low rewards for almost every actions the generator takes, which does not guide the generator towards a good direction for performance improvement, resulting in inferior performance at the beginning. However, as the training goes on, the generator and discriminator can gradually learn from each other and finally converge, which shows the stability of AICM. 

\subsubsection{Ablation Study on Training Strategy}

\begin{figure}[t]
    \centering
    \includegraphics[width=0.23\textwidth]{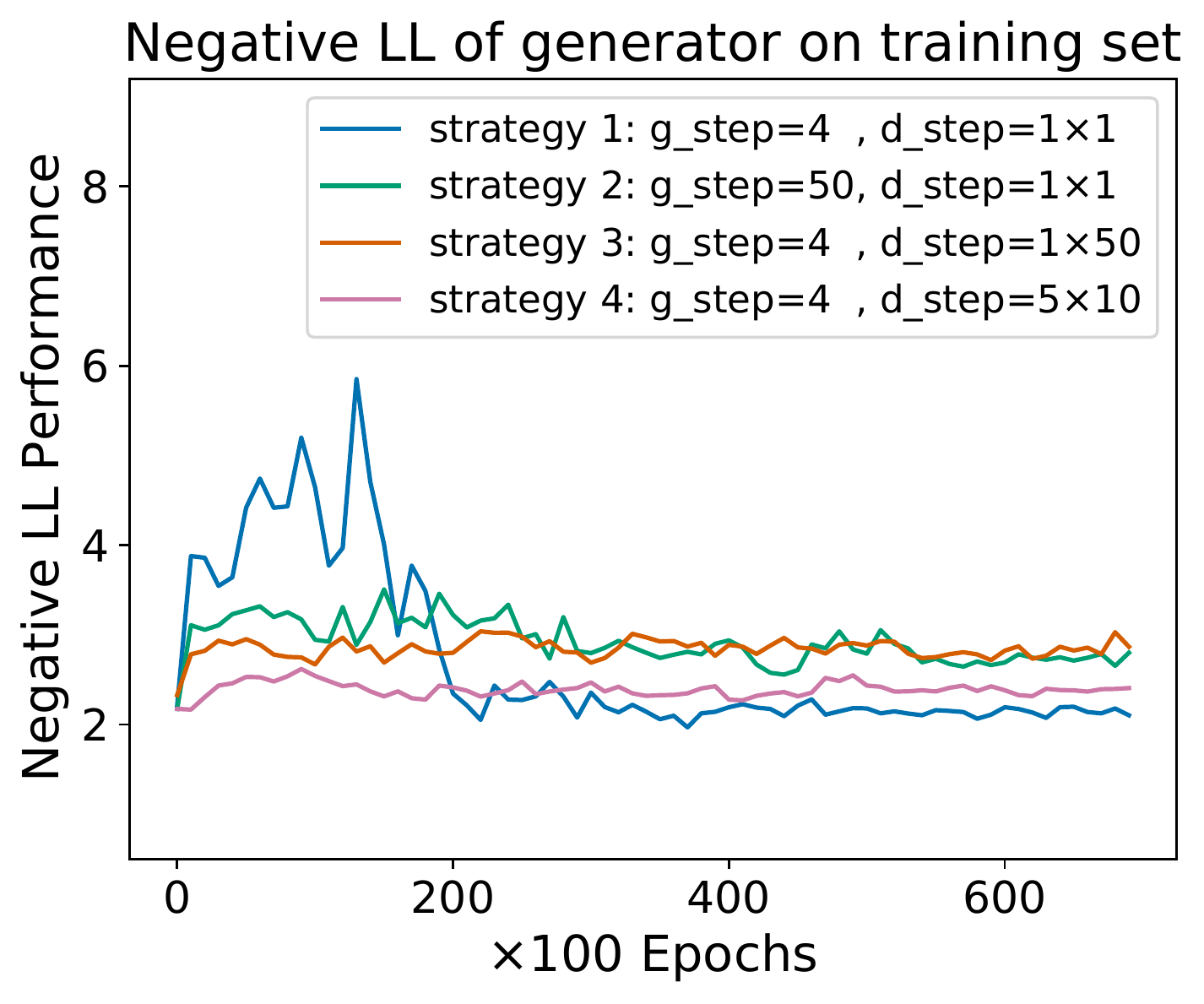}
    \includegraphics[width=0.24\textwidth]{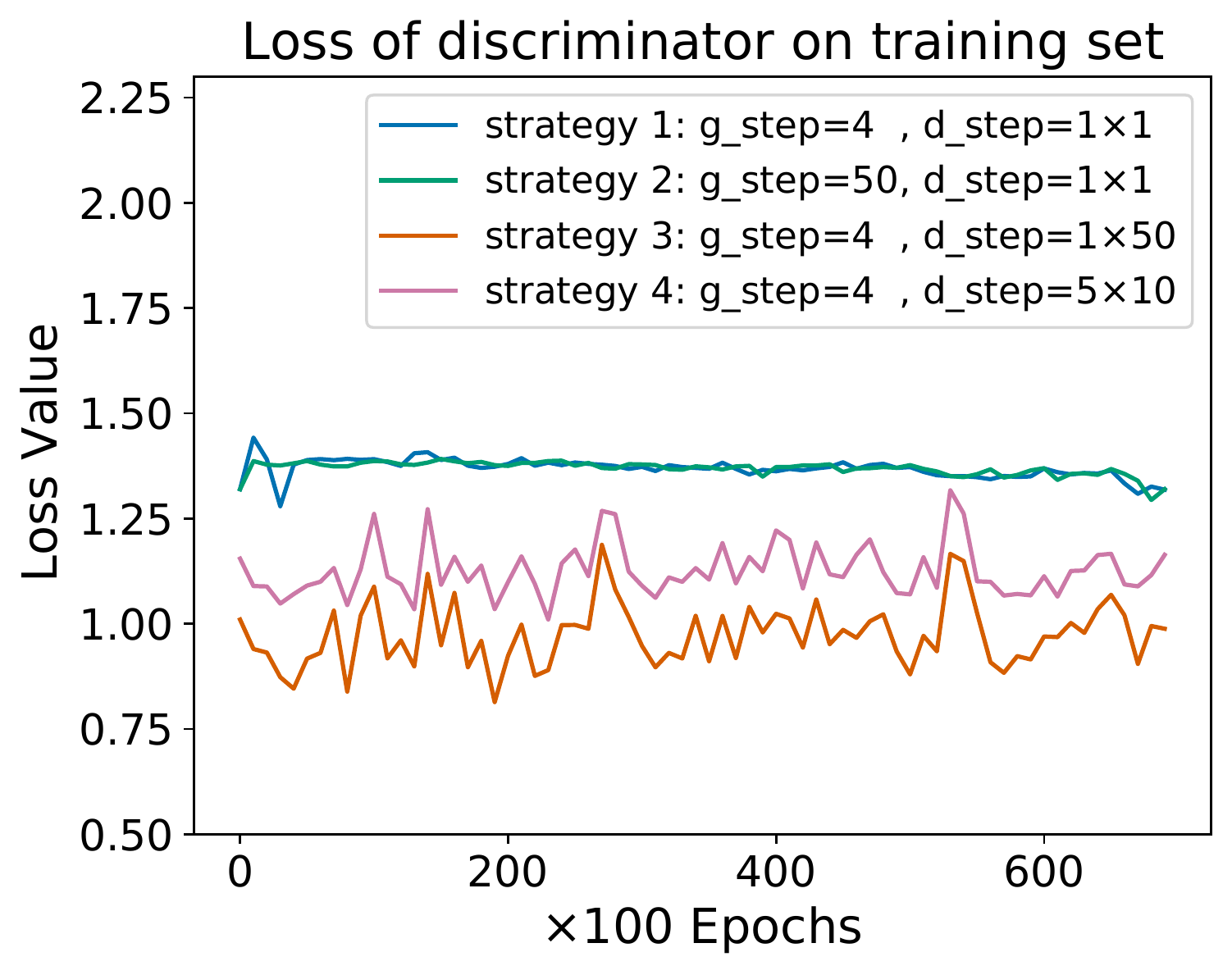}
    \caption{The comparison of negative LL performance during training period for the generator and discriminator w.r.t. different $g\_step$-$d\_step$ configuration. (Best viewed in color.)}
    \label{fig:ablation on G D steps}
\end{figure}

In our experiments, we find that the stability of AICM highly depends on training strategies. More specifically, hyper-parameters $g\_step$ and $d\_step$ have a large effect on the performance of AICM. Figure~\ref{fig:ablation on G D steps} shows the effect of these two parameters. Suppose we set $g\_step=k$ and $d\_step=m\times n$. Then, in each epoch, we train the generator for $k$ times, and use the trained generator to generate $m$ synthetic trajectories. For each trajectory $\tau$, the discriminator is trained for $n$ times, resulting in total $m\times n$ updates for the discriminator in each epoch. From Figure~\ref{fig:ablation on G D steps}, we can obtain the following observations:

\begin{itemize}[leftmargin=15pt]
    \item[(1)] Strategy 1, which is adopted in our experiments above, achieves the best performance. As the generator performs the best, the loss of the discriminator is higher than that of other strategies. In addition, the fluctuation at the beginning is caused by adversarial updates between generator and discriminator, which has been explained in Section~\ref{sec:ablation/pretrain}.
    \item[(2)] In strategy 2, $g\_step$ is much larger than $d\_step$, which leads to training the generator many times before updating the discriminator once. This strategy results in a fast convergence of the generator. However, in this case, the generator improves so quickly, that the discriminator cannot get fully trained and thus provides a misleading signal gradually. That is why strategy 2 leads to worse performance in AICM than strategy 1.
    \item[(3)] In strategy 3, the discriminator is sufficiently trained in each epoch. A fully trained discriminator can easily distinguish the fake sequence generated by an insufficiently trained generator. Thus almost every synthetic sequence receives a low reward, which does not provide a good guidance for the generator.
    \item[(4)] Compared to strategy 3, the total number of updates for the discriminator in strategy 4 is still $50$. But in each epoch, we use the generator to generate $5$ synthetic trajectories and update the discriminator 10 times for each trajectory. 
    This alleviates the overfitting of the discriminator and provides meaningful signal to the generator.
    Thus, the negative LL performance of the generator in strategy 4 is much better compared with strategy 3.
    However, strategy 4 performs worse than strategy 1 since the discriminator is still overtrained.
\end{itemize}

From the analysis above, we conclude that AICM benefits from a proper ratio of $g\_step$ and $d\_step$, which is in line with the theorem in~\cite{goodfellow2014generative}. It is important to balance the training of the generator and discriminator. Only if the discriminator is capable of consistently differentiating real data from generated data, which should not be too simple to be distinguished, the supervised signal from discriminator can be meaningful and the whole adversarial training process can be stable and effective.

\subsubsection{Ablation Study on Discount Factor $\gamma$}

\begin{figure}[h]
    \centering
    \includegraphics[width=0.48\textwidth]{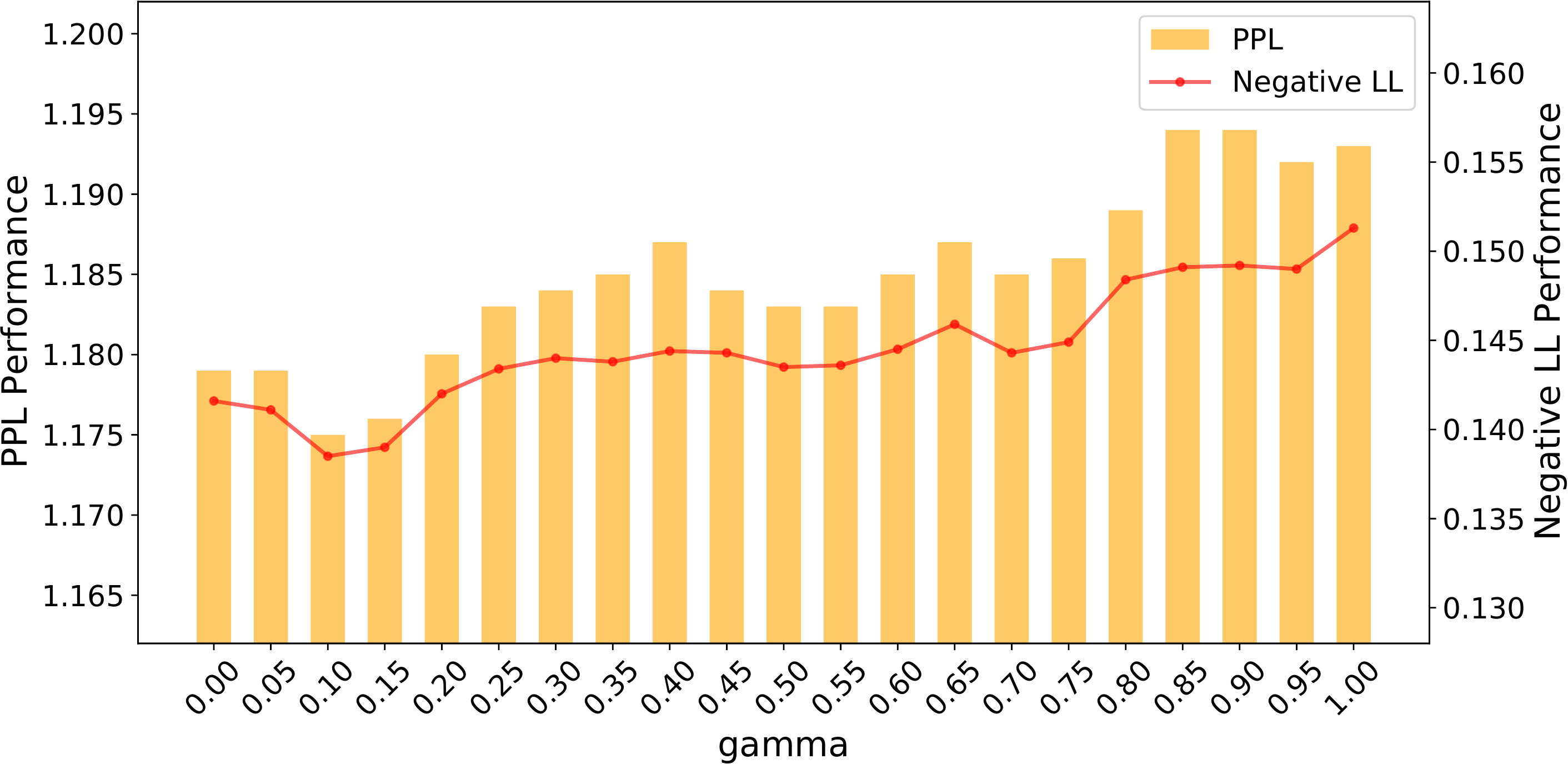}
    \caption{The comparison of PPL and negative LL performance w.r.t. different $\gamma$ values.}
    \label{fig:ablation on gamma}
\end{figure}

The discount factor $\gamma$ controls how much of the future we should look ahead to make the current decision. Typically, $\gamma$ is viewed as part of the problem. However, in practice, we need to tune this parameter to obtain the best value that is suitable for certain tasks. In Figure~\ref{fig:ablation on gamma}, we show the PPL and negative LL performance on test set w.r.t different $\gamma$ values. The best performance is obtained at $\gamma=0.1$. This is a small value, showing a large discount on future rewards. This is reasonable due to the existence of position bias. The state in the distant future often corresponds to a lower position and becomes less important. 

%% file: text/related.tex
\section{Related Work}
We first describe the prior works in click models, then we discuss the connections and distinctions between AICM and previous GAN/GAIL based user simulation models. 

\minisection{Click Models}
Traditional click models~\cite{chuklin2015click}, which are based on PGM framework, treat user behaviors as a sequence of observable and hidden events. They usually incorporate different assumptions on user behaviors to specify how documents and clicks at different positions affect each other. \citet{richardson2007predicting} proposed the \textit{examination hypothesis}, under which the probability of click are decomposed into the examination probability and the document relevance. 
Different click models study examination probability differently. The simplest click model that follows the examination hypothesis is the position-based model (PBM)~\cite{craswell2008experimental}, which assumes that the examination probability only relates to the displayed positions. \citet{craswell2008experimental} proposed the cascade model (CM) by assuming that users sequentially scan each document in the list until the first click. CM can only handle query sessions with exactly one click. On the basis of CM, user browsing model (UBM), dynamic Bayesian network (DBN), dependent click model (DCM), and click chain model (CCM) have been proposed to overcome this limitation. 

To get better expressive power and flexible dependencies, NN-based approaches have been proposed. The neural click model (NCM)~\cite{borisov2016neural} is the first attempt to apply neural networks to click models. NCM represents user behaviors as a sequence of hidden states instead of binary events. The following neural network based approaches also adopt this distributed representation framework. The click sequence model (CSM)~\cite{borisov2018click} incorporates an encoder-decoder architecture, where the encoder computes contextual embeddings of the documents and the decoder predicts the position sequence of the clicked documents. The contextual-aware click model (CACM)~\cite{chen2020context} takes the session-level information into consideration and separates the modeling of relevance and examination. Such methods suffer from exposure bias and inferior estimation, which is successfully alleviated in AICM by dynamic modeling and adversarial training. 

\minisection{GAN/GAIL based User Simulation}
The framework of GAN/GAIL has been successfully adopted for user simulation in many previous works~\cite{shi2019virtual, chen2019generative, bai2019model}. These works are mostly built as simulators to enhance RL-based recommendation agents. These works model cross-page interactions, which differs from AICM in problem definition. 
VirtualTaobao~\cite{shi2019virtual} models state transitions as turning to next page or switching to another user, and users' actions are simply defined as her interactions to the whole page, i.e., buying, leaving, or turning page. 
GAN-CDQN~\cite{chen2019generative} models state transitions as turning to the next page, and define users' actions as picking an item (or not pick) from the $k$ item set regardless of the order of item lists. Similar as above, IRecGAN~\cite{bai2019model} also models state transitions as turning to the next page and users' actions are defined as a click or not on a ranked item of the list. 

Such a modeling is restrictive and many important details of user behaviors might be lost, e.g., the rank of clicked items, the context of clicked items, even which item is clicked (in VirtualTaobao and IRecGAN). 
Also, it cannot deal with multiple clicks, which are common in real-world applications. 
Moreover, none of the mentioned models can be used to evaluate a ranking function since they simply ignore the order of the ranked list. 
In AICM, we focus on users' interaction with a ranked list and model a fine-grained user behavior within the ranked list, which provides useful information for both the training and evaluation of a ranking function.

\section{Conclusion}
In this work, we propose a novel learning paradigm for click models based on the imitation learning framework. We model users' interaction with a ranked list as a sequential decision-making process instead of one-step prediction, and learn a multi-step click policy from users' click logs as expert demonstrations. We base the users' current state on previous predictions and optimize for a long-term objective rather than a short-sighted one-step loss. With adversarial training, we learn a stable distribution which generalizes well across different ranked list distributions. Also, we explicitly build a reward function, which recovers users' intrinsic utility and underlying intentions. 
Theoretical analysis shows that our solution is capable of reducing the exposure bias from $O(T^2)$ to $O(T)$. Empirical studies on a real-world web search dataset demonstrate the effectiveness of our solution from different aspects. For future work of research, we will utilize AICM in the offline evaluation and optimization for a ranking function.

%% file: text/Appendix.tex
\begin{table}[b]
    \centering
    \caption{Computation Time.}
    \label{tab:computation_time}
    \vspace{-9pt}
    \begin{tabular}{c c c c}
    \toprule
     & NCM & CACM & AICM \\
    \midrule
    pretrain & 0 & 2h3min & 24min \\
    train & 1h40min & 12h34min & 3h30min \\
    total & 1h40min & 14h37min & 3h54min \\
    \bottomrule
    \end{tabular}
    \vspace{-9pt}
\end{table}
\section{Detailed proofs}
The proof in this section can be seen as a finite version of \cite{xu2019value}.
Here we choose an easier way compared to the original proof in the first theorem due to the introduction of the specific state transition in click model, which are shown in Eq.~\eqref{eq:thrm1/2} and Eq.~\eqref{eq:thrm1/3}.
\subsection{Proof for Theorem~\ref{thrm:bc}}
\label{proof:bc}
\begin{proof}
Note that
\begin{equation}
\small{
\begin{aligned}
    & |J(\pi) - J(\pi_E)| \\
    =& |\sum\nolimits_{t=0}^{T}  \gamma^t  \sum\nolimits_{s_t,a_t} (P_\pi(s_t, a_t) - P_{\pi_E}(s_t, a_t))R(s_t, a_t)|\\
    \le& \sum\nolimits_{t=0}^{T} \gamma^t  |\sum\nolimits_{s_t,a_t} (P_\pi(s_t, a_t) - P_{\pi_E}(s_t, a_t))R(s_t, a_t))|\\
    = & \sum\nolimits_{t=0}^{T} \gamma^t |\sum\nolimits_{s_t,a_t} (P_\pi(s_t) \pi(a_t|s_t) - P_{\pi_E}(s_t)\pi_E(a_t|s_t))R(s_t, a_t)|\\
    = & \sum\nolimits_{t=0}^{T} \gamma^t  |\sum\nolimits_{s_t,a_t} (P_\pi(s_t) \pi(a_t|s_t) - P_{\pi_E}(s_t) \pi(a_t|s_t)\\
    & + P_{\pi_E}(s_t) \pi(a_t|s_t) - P_{\pi_E}(s_t)\pi_E(a_t|s_t))R(s_t, a_t)|\\
    \leq &  \sum\nolimits_{t=0}^{T} \gamma^t R_{\max}\bigg( \sum\nolimits_{s_t,a_t} |P_\pi(s_t) -P_{\pi_E}(s_t)| \pi(a_t|s_t)\\
    & + \sum\nolimits_{s_t,a_t} |\pi(a_t|s_t) - \pi_E(a_t|s_t)|P_{\pi_E}(s_t)\bigg)\\
    \leq &  \sum\nolimits_{t=0}^{T} \gamma^t R_{\max}\bigg( \sum\nolimits_{s_t}|P_\pi(s_t) -P_{\pi_E}(s_t)| \sum\nolimits_{a_t} \pi(a_t|s_t) \\ 
    & + \mathbb{E}_{s_t\sim \pi_E} [\sum\nolimits_{a_t}|\pi(a_t|s_t) - \pi_E(a_t|s_t)|]\bigg)\,.\\
\end{aligned}
}
\label{eq:thrm1/1}
\end{equation}
The first absolute term can be further decomposed as
\begin{equation}
\small{
\begin{aligned}
    & \sum\nolimits_{s_t}|P_\pi(s_t) -P_{\pi_E}(s_t)| \\
    = &  \sum\nolimits_{s_{t-1}, a_{t-1}, d_t} |P_\pi(s_{t-1}, a_{t-1}, d_t) - P_{\pi_E}(s_{t-1}, a_{t-1}, d_t)|\\
    = &  \sum\nolimits_{s_{t-1}, a_{t-1}, d_t} |P_\pi(s_{t-1}, a_{t-1}) - P_{\pi_E}(s_{t-1}, a_{t-1})|P(d_t|s_{t-1}, a_{t-1})\\
    = &  \sum\nolimits_{s_{t-1}, a_{t-1}} |P_\pi(s_{t-1}, a_{t-1})-P_{\pi_E}(s_{t-1}, a_{t-1})| \sum\nolimits_{d_t} P(d_t|s_{t-1}, a_{t-1}))\\
    = &  \sum\nolimits_{s_{t-1}, a_{t-1}} |P_\pi(s_{t-1}, a_{t-1}) -P_{\pi_E}(s_{t-1}, a_{t-1})|\\
    = &  \sum\nolimits_{s_{t-1}, a_{t-1}} |P_\pi(s_{t-1})\pi(a_{t-1}|s_{t-1}) - P_{\pi_E}(s_{t-1}) \pi(a_{t-1}|s_{t-1})\\
    & + P_{\pi_E}(s_{t-1}) \pi(a_t|s_{t-1}) -P_{\pi_E}(s_{t-1})\pi_E(a_{t-1}|s_{t-1})|\\
    \leq &  \sum\nolimits_{s_{t-1}, a_{t-1}} |P_\pi(s_{t-1}) - P_{\pi_E}(s_{t-1})| \pi(a_{t-1}|s_{t-1})\\
    & + |P_{\pi_E}(s_{t-1}) -P_{\pi_E}(s_{t-1})|\pi_E(a_{t-1}|s_{t-1})\\
    = & \sum\nolimits_{s_{t-1}} |P_\pi(s_{t-1}) -P_{\pi_E}(s_{t-1})|\sum\nolimits_{a_{t-1}} \pi(a_{t-1}|s_{t-1})) \\
    & + \sum\nolimits_{s_{t-1}, a_{t-1}} |\pi(a_{t-1}|s_{t-1}) - \pi_E(a_{t-1}|s_{t-1})|P_{\pi_E}(s_{t-1})\\
    = & \sum\nolimits_{s_{0}} |P_\pi(s_{0}) -P_{\pi_E}(s_{0})| \\
    & + \sum\nolimits_{k=1}^{t-1} \sum\nolimits_{s_{k}, a_{k}} |\pi(a_{k}|s_{k}) - \pi_E(a_{k}|s_{k})|P_{\pi_E}(s_{k})\\
    = & \sum\nolimits_{k=1}^{t-1} \mathbb{E}_{s_k\sim \pi_E} [\sum\nolimits_{a_k} |\pi(a_{k}|s_{k}) - \pi_E(a_{k}|s_{k})|].
\end{aligned}
}
\label{eq:thrm1/2}
\end{equation}

Combining Eq.~\eqref{eq:thrm1/1} and Eq.~\eqref{eq:thrm1/2}, we have
\begin{equation}
\small{
\begin{aligned}
    &|J(\pi) - J(\pi_E)| \\
    \leq &  \sum\nolimits_{t=0}^{T} \gamma^t R_{max} \sum\nolimits_{s_t}|P_\pi(s_t) -P_{\pi_E}(s_t)| \\ 
    & + \mathbb{E}_{s_t\sim \pi_E} [\sum\nolimits_{a_t}|\pi(a_t|s_t) - \pi_E(a_t|s_t)|]\\
    = & \sum\nolimits_{t=0}^{T} \gamma^t R_{max} \sum\nolimits_{k=1}^{t}  \mathbb{E}_{s_k\sim \pi_E} [\sum\nolimits_{a_k} |\pi(a_{k}|s_{k}) - \pi_E(a_{k}|s_{k})|] \\
    = & 2 \sum\nolimits_{t=0}^{T} \gamma^t R_{max} \sum\nolimits_{k=1}^{t}  \mathbb{E}_{s_k\sim \pi_E} [D_{TV}(\pi_E(\cdot|s_k), \pi(\cdot|s_k))]\\
    \leq & 2 \sum\nolimits_{t=0}^{T}  \gamma^t R_{max} \sum\nolimits_{k=1}^{t} \mathbb{E}_{s_k\sim \pi_E} [\sqrt{\frac{1}{2} D_{KL} (\pi_E(\cdot|s_k), \pi(\cdot|s_k))}] \\
    \leq & \sqrt{2} \sum\nolimits_{t=0}^{T}  t \gamma^t R_{max} \sqrt{\epsilon_{bc}}\\
    \leq & \sqrt{2} T(T+1)  R_{max} \sqrt{\epsilon_{bc}}
\end{aligned}
}
\label{eq:thrm1/3}
\end{equation}
where total variation between two distributions is defined as 
\begin{equation}
D_{TV}(P, Q) = \frac{1}{2} \|P-Q\|_{1} = \frac{1}{2} \sum_x |P(x)-Q(x)|. 
\end{equation} 
\end{proof}

\subsection{Proof for Theorem~\ref{thrm:gail}}
\label{proof:gail}
\begin{proof}
\begin{equation}
\small{
\begin{aligned}
    & |J(\pi) - J(\pi_E)| \\
    = & \sum_{t=0}^{T} |  \gamma^t \sum\nolimits_{s_t,a_t} P_\pi(s_t, a_t) - \sum_{t=0}^{T}  \gamma^t P_{\pi_E}(s_t, a_t)R(s_t, a_t)|\\
    \leq & 2  \sum_{t=0}^{T} \sum\nolimits_{s_t,a_t} |  \gamma^t P_\pi(s_t, a_t) -  \gamma^t P_{\pi_E}(s_t, a_t)| R_{max}\\
    = & \frac{2(1-\gamma^{T+1})}{1-\gamma} D_{TV}( \rho_{\pi}, \rho_{\pi_E}) R_{max}\\
    = & \frac{2(1-\gamma^{T+1})}{1-\gamma} \sqrt{2(D_{TV}^2(\rho_{\pi},\frac{\rho_{\pi}+\rho_{\pi_E}}{2})+ D_{TV}^2(\rho_{\pi},\frac{\rho_{\pi}+\rho_{\pi_E}}{2}))} R_{max}\\
    \leq & \frac{2(1-\gamma^{T+1})}{1-\gamma} \sqrt{2\times \frac{1}{2}(D_{KL}(\rho_{\pi},\frac{\rho_{\pi}+\rho_{\pi_E}}{2})+ D_{KL}(\rho_{\pi}, \frac{\rho_{\pi}+\rho_{\pi_E}}{2}))} R_{max}\\
    = & \frac{2(1-\gamma^{T+1})}{1-\gamma} \sqrt{2 D_{JS}(\rho_{\pi}, \rho_{\pi_E})} R_{max}\\
    = & \frac{2\sqrt{2}(1-\gamma^{T+1})}{1-\gamma} \sqrt{\epsilon_{ga} } R_{max}\\
    \leq & 2 \sqrt{2}  R_{max} (T+1) \sqrt{\epsilon_{ga} }\,.\\
\end{aligned}
}
\end{equation}
\end{proof}

\section{Computation Time}

Compared to the state-of-art model CACM, AICM actually has simpler model structure and consequently, lower computational complexity. AICM uses two separate RNNs and adversarially trains them, while CACM uses four encoders, which are respectively 3 RNNs with attention mechanism and 1 MLP, and jointly updates them. We report the time spent on the whole training phases of NCM, CACM and AICM with the TianGong ST dataset under the same GPU environment in Table~\ref{tab:computation_time}.